# Generating Circularly Polarized and Chiral Cathodoluminescence from a Single Plasmonic Nanoantenna

Stan de Peinder[1*], Hollie Marks[1], Quynh Nguyen[1], Álvaro Rodríguez Echarri[1] and Albert Polman[1]

Center for Nanophotonics, NWO Institute AMOLF, Science Park 104, 1098XG Amsterdam, The Netherlands[1]

## Abstract

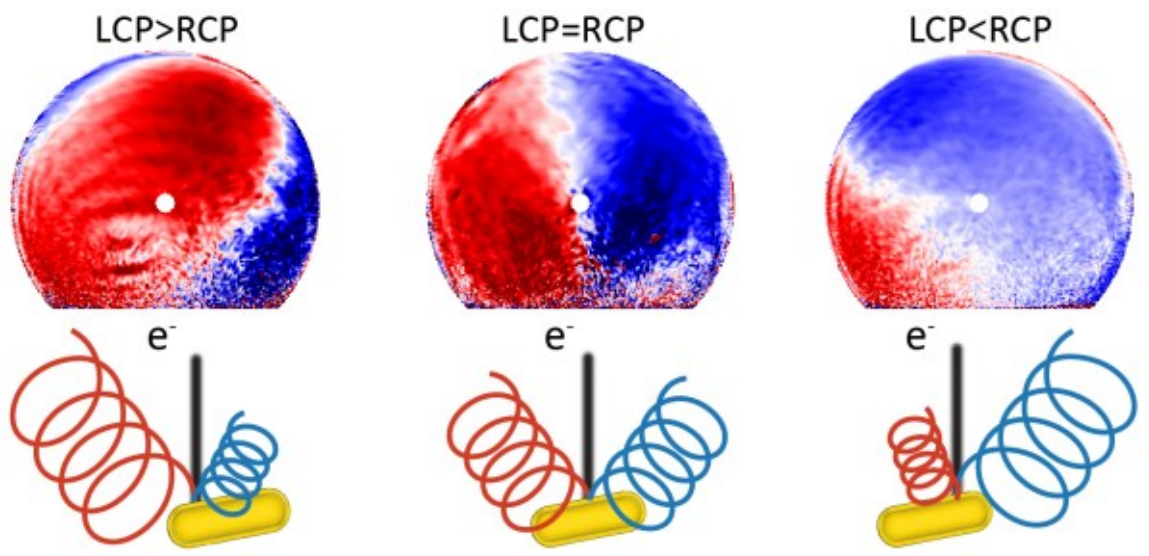

The generation and control of circularly polarized and chiral light is important for a wide range of applications, from sensing to optical communication and catalysis. Generally, the generation of chiral light emission from plasmonic nanostructures relies on the use of nanoantennas with predefined chiral shape, which requires complex nanofabrication. In this work, we show how localized electron excitation controls the generation of circularly polarized- and chiral cathodoluminescence from a single achiral plasmonic nanoantenna. We exploit the $\pi/2$ phase difference between the transverse and longitudinal components of the electric field that is carried by 30 keV electrons, to coherently excite degenerate dipolar modes of which the far-field radiation interferes in the far field, thereby creating a strong optical spin-Hall effect. Using angle-resolved cathodoluminescence polarimetry we measure the angle-resolved Stokes parameters for selected positions of the electron beam near the nanoparticle and show how the precise location of the electron beam controls the generation of the circularly polarized and chiral electric field components. Our findings represent an elemental form of generation of circularly polarized light, by synthesizing the circularly rotating electric fields characteristic for CPL from interference of orthogonal dipolar modes with controlled phase lag.

## Introduction

There is a growing interest in controlling the generation of circularly polarized light (CPL) due to its applicability in sensing, optical communication, catalysis, 3D displays, encryption, quantum communication, and more.[1–6] The left- and right-handed polarization can encode orthogonal information channels, selectively sense molecular enantiomers, and selectively drive molecular reactions, to mention a few examples.[7–9]

Noble metal plasmonic optical nanoantennas offer a tunable platform for the generation of CPL at the nanoscale due to their localized surface plasmonic resonances (LSPRs) that strongly enhance the electromagnetic near field, and that are tunable over a wide spectral range by controlling nanoparticle shape and size.[10,11] One approach to generate CPL is to excite plasmonic nanostructures with linearly polarized light with the polarization axis rotated relative to the symmetry axis of the nanostructure, exciting orthogonal multi-polar plasmonic modes that are out of phase in such a way that they interfere in the far field and create CPL.[12] Controlling the handedness and directionality of CPL has also been demonstrated using nanopatterned 2D chiral nanostructures that scatter polarization components with a phase offset such that they create CPL.[13,14] A special case is the generation of CPL where one unique handedness dominates the far field (chiral emission). [15–18] The generation of chiral emission at the nanoscale often relies on intrinsically chiral objects such as nanohelices or colloidally synthesized chiral nanoparticles with well-defined handedness.[16,19,20] All these structures have significant complexity, and in most cases involve excitation of multiple plasmonic modes. This raises the question: what is the simplest nanophotonic structure from with chiral light can be generated?

To investigate this, we use high-energy electron excitation, which is uniquely suited as a localized excitation source of plasmonic structures.[21–23] The time-varying electric field carried by a swift electron (1-100 keV) represents a broad spectral range of excitation (0-30 keV),[24] and when applied in a Scanning or Transmission Electron Microscope (SEM, TEM) provides nanoscale spatial excitation resolution. This enables the broadband excitation of cathodoluminescence (CL) of plasmonic nanostructures into the far field, where its spectrum and polarization can be detected and analyzed.[25,26] The generation of CPL and chiral CL emission in SEM or TEM has been shown in Si nanospheres.[27] Au nanorods, [28] Au nanocrescents[29], and V-shaped Al nanoantennas. [30]. Gold nanoantenna arrays of which the modes interfere in the far field were also shown to emit chiral CL emission.[31,32] In all these works, however, chiral emission was due to multipolar interference.

Here we demonstrate how coherent superposition of the emission of pure electric dipole modes, the most elemental nanophotonic building blocks, creates chirality. The chiral light is synthesized by taking advantage of a $\pi/2$ phase difference of degenerate orthogonal dipole modes that is controlled by the phase difference between transverse and longitudinal electric fields carried by the electron. We show how we create strongly antisymmetric LCP and RCP CL emission in momentum space (optical spin-Hall effect, OSHE) with a directionality controlled by the electron beam position. By breaking the excitation symmetry we then create strongly chiral emission, with the handedness and propagation direction controlled by the placement of the electron beam.

## Results and discussion

### CL spectroscopy and Angle-resolved polarimetry

Gold nanorods with minor and major axes 50×110 nm were colloidally synthesized, see Methods; a representative TEM image is shown in Figure 1a. CL spectroscopy is carried out in the SEM at 30 keV in an aloof geometry with the electron beam placed 5 nm from the surface. Figure 1b (solid lines) shows the resonant transverse (535 nm) and longitudinal (650 nm) dipolar plasmon modes for the impact parameters indicated by the colored dots in the schematic. The surface charge distributions for the resonant modes are shown in Figure S1 (see SI). The dashed lines in Figure 1b show boundary element method (BEM) simulations for the two excitation conditions (see Methods)[33] they match the resonance wavelengths very well.

### Angle-resolved CL polarimetry

To investigate the polarization characteristics of the CL emission, we use angle-resolved CL polarimetry in the SEM, see the schematic in Figure 1c and Methods.

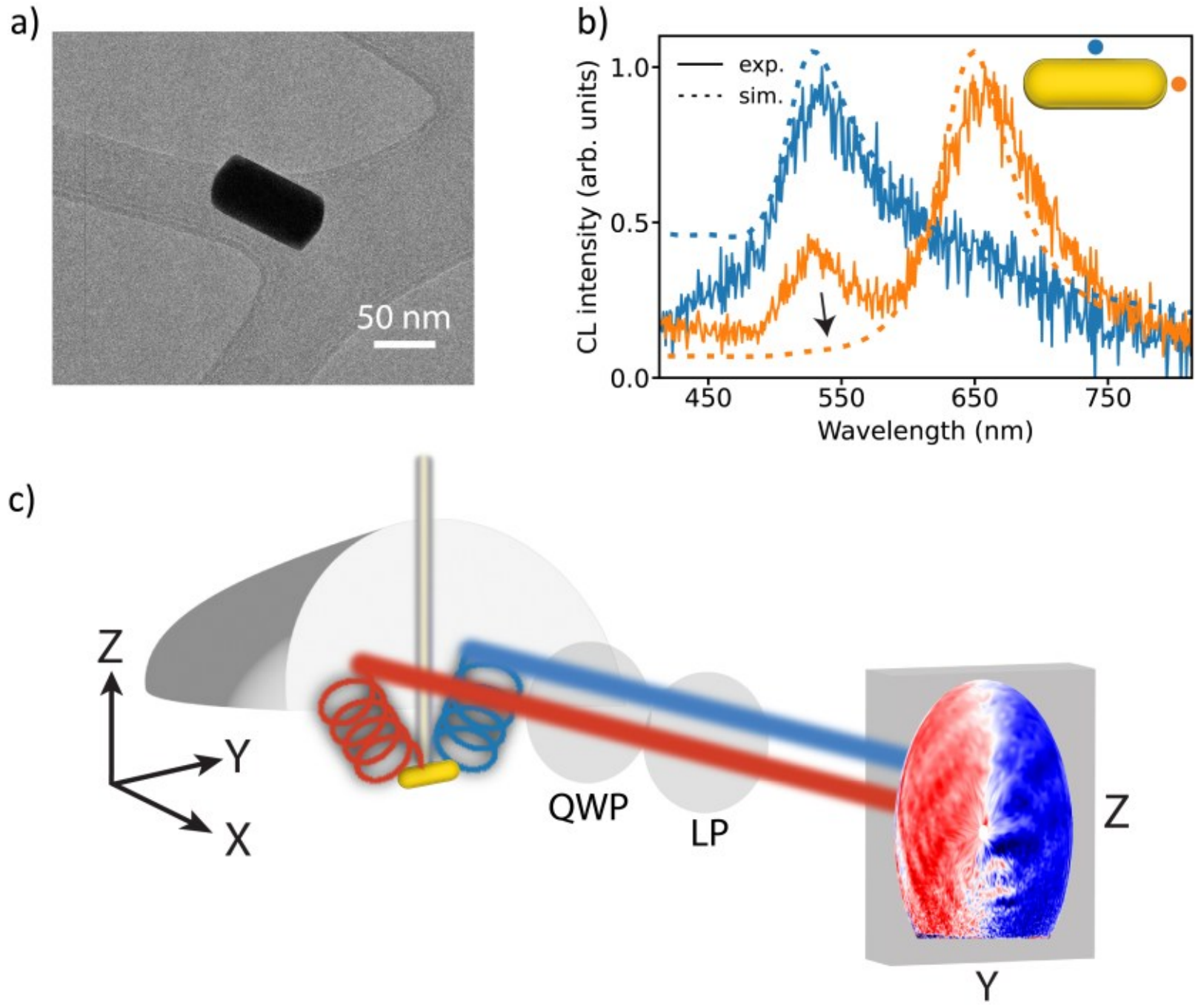

Figure 1. a) TEM image of a representative Au nanorod used in this work. b) Measured (solid) and simulated (dashed) CL spectra obtained by aloof excitation of a Au nanorod by a 30 keV electron, at the impact parameters indicated by the colored dots in the schematic. The black arrow points to the transversal resonance contribution in the simulated results. c) Schematic of the angle-resolved polarimetry configuration with a quarter-wave plate (QWP) and linear polarizer (LP) in the optical path, and the definition of the coordinate system used in this work.

Rotating a quarter-wave plate and a linear polarizer allows for the detection of transmitted intensities at six different orientations of the polarizer components: vertical, horizontal, negative diagonal (45°), positive diagonal (135°), left-handed circular (LHC) and right-handed circular (RHC). Measuring the angle-resolved intensity distribution of these six polarization components allows for the reconstruction of the four Stokes parameters (S0 to S3) which are defined in Methods (eqn. 7). The Stokes parameters represent the complete polarization information of the collected CL and can be used to express any desired polarization metric. Reconstructing the Stokes parameters requires correcting the acquired data with the Mueller matrix of the optical system, as the Al parabolic mirror significantly alters the polarization of the collected light. For further details we refer the reader to the work by Osorio et al.[26]

## CPL generation by controlling transversal dipolar modal interference

Figure 2 shows angle-resolved CL polarimetry results for aloof excitation next to the nanorod at two different positions. In Figure 2a we show the normalized Stokes parameter S3/S0, which reflects the difference between the left-and right-handed circularly-polarized light intensities (see Methods), for excitation at the position of the blue dot. The corresponding angle-resolved CL intensity distributions of the emitted LCP and RCP intensities are shown in Figures 2b, and c. A high-intensity LCP emission lobe is mostly directed to the quadrant corresponding to $+k_x$ and $-k_y$ wavevectors, while RCP emission is most strongly emitted into the quadrant corresponding to $-k_x$ and $-k_y$ wavevectors. Similar data are observed for excitation for the electron trajectory at the mirror-symmetric position relative to the $\sigma_{yz}$ mirror plane (Figure S4), but now mirror imaged, confirming the symmetric nature of the CL emission contributions. We note that in Figure 2a the axis dividing the S3/S0 intensity lobes is rotated away slightly from $k_x$=0, which we attribute to the slight misalignment of the nanorod position with respect to the optical axis of the light collection system in that measurement.

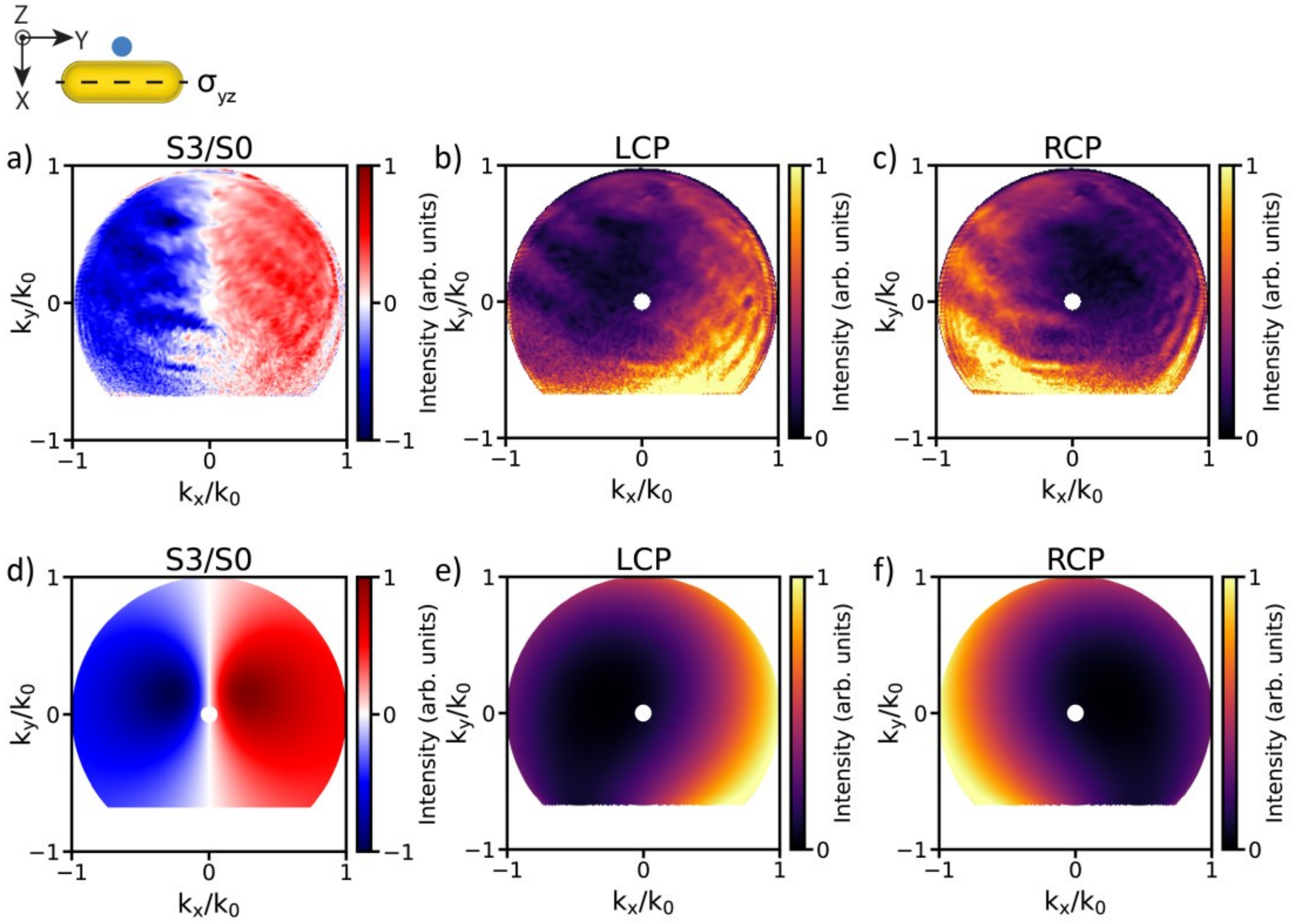

Figure 2. Experimental and calculated angle-resolved CL data for 30 keV aloof electron excitation of the Au nanorod (5 nm from the surface) at the indicated positions (blue dot), using a bandpass filter at 550±25nm. (a,d) Normalized S3 Stokes parameter, (b,e) left-handed circularly polarized (LCP) and (c,f) right-handed circularly polarized (RCP) CL intensity.

The data show the optical spin-Hall effect (OSHE), in which the handedness of the angular momentum of light is coupled to its propagation direction. CPL is created by the interference between two orthogonal dipolar modes that are excited with a phase delay that is inherent to the electron excitation mechanism, as we will show next. The control of the scattering phase of the individual dipoles composes a circularly rotating electromagnetic far field that creates CPL with different handedness for the two directions perpendicular to the symmetry plane of the excitation geometry, i.e. along the longitudinal axis of the particle.

The electric field $\boldsymbol{E}(\boldsymbol{r},\omega)$ at a position $\boldsymbol{r}$, at frequency of $\omega$ of a swift electron travelling in vacuum at a velocity $v$ along the z axis is given by

$$\boldsymbol{E}(\boldsymbol{r},\omega) = \left(\frac{2e\omega}{v^2\gamma\epsilon_d}\right) e^{i\omega z/v}\left[\frac{i}{\gamma}\boldsymbol{K}_0\left(\frac{\omega R}{v\gamma}\right)\hat{\boldsymbol{z}} - \boldsymbol{K}_1\left(\frac{\omega R}{v\gamma}\right)\hat{\boldsymbol{R}}\right] \quad (1)$$

with $\gamma$ the Lorentz contraction factor, and $K_0$ and $K_1$ the modified Bessel functions of the second kind.[34] The first term in the square brackets describes the longitudinal field component $E_z$, that is aligned with the trajectory of the electron, while the second term describes the radial field component $E_R$. With the first term purely imaginary and the second real, the two field components are in quadrature; they have a phase difference of exactly π/2.

For the small Au nanorods used in this work, CL emission is dominated by dipolar contributions, as the quadrupolar modes do not radiate efficiently (see Figure 1b). Therefore, we use the electric dipole approximation and define the dipole moment induced in a gold nanorod after electron excitation

$$\boldsymbol{p}(\omega) = \boldsymbol{\alpha}(\omega)\cdot\boldsymbol{E}(\hat{\boldsymbol{R}},\omega) \quad (2)$$

with $\boldsymbol{\alpha}(\omega)$ the frequency-dependent complex polarizability tensor of the nanorod, $\hat{\boldsymbol{R}}$ the impact parameter, and $\boldsymbol{E}$ the electric field associated with the swift electron. Using the boundary-element method (BEM)[35], the induced electric near-field is calculated for a swift electron at an impact parameter of 5 nm distance perpendicular from the nanorod surface; details are found in the Methods section. We then derive the induced surface charge distribution $\sigma_{\mathrm{ind}}(\omega)$ from the discontinuity equation

$$\sigma_{\mathrm{ind}}(\omega) = -\varepsilon_d\left[\boldsymbol{E}_{\mathrm{out}}^{\mathrm{ind},\perp}(\omega) - \boldsymbol{E}_{\mathrm{in}}^{\mathrm{ind},\perp}(\omega)\right], \quad (3)$$

with $\boldsymbol{E}_{\mathrm{out}}^{\mathrm{ind},\perp}(\omega)$ and $\boldsymbol{E}_{\mathrm{in}}^{\mathrm{ind},\perp}(\omega)$ the induced electric field components normal to the surface at the outer and inner side of the surface, respectively, and $\varepsilon_d$ the dielectric constant of the environment. From the induced surface charge distribution, the frequency dependent dipole moment $\boldsymbol{p}(\omega)$ can be calculated from

$$\boldsymbol{p}(\omega) = \sum_i \boldsymbol{r}_i\,\sigma_i(\omega)A_i, \quad (4)$$

with $\mathbf{r}_i = (x_i, y_i, z_i)$ the position of each discrete surface element with surface charge $\sigma_i$ and area $A_i$, from which we calculate the electric field that is radiated to the far field:

$$\boldsymbol{E}_{far}(\boldsymbol{r},\omega) = \frac{k^2}{4\pi\varepsilon_d}\frac{e^{ikr}}{r}(\boldsymbol{I} - \hat{\boldsymbol{r}}\otimes\hat{\boldsymbol{r}})\cdot\boldsymbol{P}(\omega). \quad (5)$$

Combining equations 3-5, we obtain the radiated CL far field, where the constant pre-factors and propagation factor are omitted for clarity:

$$\mathbf{f}_{\hat{\mathbf{r}}}^{CL}(\hat{\boldsymbol{R}},\omega) = k^2(\boldsymbol{I} - \hat{\boldsymbol{r}}\otimes\hat{\boldsymbol{r}})\cdot\boldsymbol{\alpha}(\omega)\cdot\boldsymbol{E}(\boldsymbol{R},\omega). \quad (6)$$

Here $k$ is the wavenumber, and $\mathbf{I} - \hat{\mathbf{r}}\otimes\hat{\mathbf{r}}$ the operator that calculates the component of the induced dipole that radiates to the far field at position $\hat{\boldsymbol{r}}$. By projecting $\mathbf{f}_{\hat{\mathbf{r}}}^{CL}$ onto the $s$ and $p$ polarization basis, the Stokes parameters in the sample plane are calculated. Subsequently, the Mueller matrix of the optical system is used to project the angle-resolved polarimetry results to the detector plane for comparison with experimental results (see Methods).

For best correspondence between the calculations and the measured angle-resolved S3/S0 data presented in Figure 2, the strength of the $p_z$ dipole was increased by a factor of 4.15 with respect to the $p_x$ dipole, with a phase shift of +0.15π (see Figure 3a,b). As can be seen, we find good quantitative agreement between the experimental and simulated results with strong beaming of CPL of different handedness in complementary directions. The discrepancy in the strength of $p_z$ is attributed to the presence of a substrate in the experiments, which breaks the symmetry of the gold nanorod, hereby changing the boundary conditions, and the presence of a thin carbon layer that forms at the surface of the metal nanorods during electron irradiation. Neither of these effects were taken into account in our simulation results. Due to the enhanced dielectric constant compared to vacuum both effects can lead to the out-of-plane dipole being excited more efficiently. The full angle-resolved polarimetry dataset which is used to find the effective amplitude and phase shift corrections is presented in the SI (Section 3).

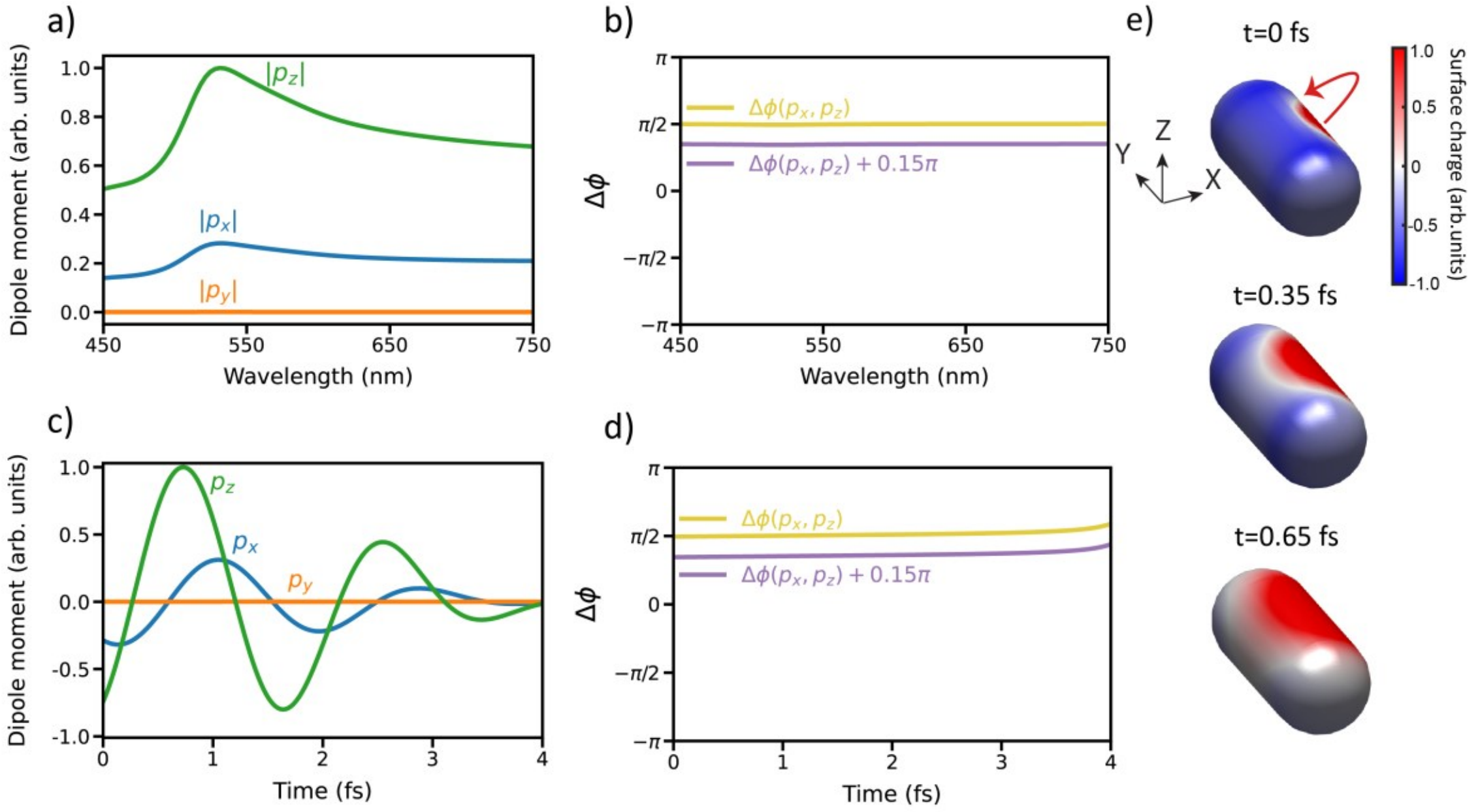

Figure 3. a) BEM-calculated induced dipole moment spectra for the three dipole orientations for 30 keV aloof electron excitation of the Au nanorod at the side (blue dot in Figure 2, 5 nm from the surface). Data for $p_z$ are scaled to match the measured S3/S0 data. b) Spectra of phase difference between $p_x$ and $p_z$, as calculated (yellow) and corrected to match the measured S3/S0 data (purple). c) Time evolution of dipole moments $p_x$, $p_y$, $p_z$, with t=0 is defined as the point in time where the electron is closest to the nanostructure. d) Time evolution of the phase difference between $p_x$ and $p_z$ as calculated with BEM (yellow) and corrected to match the measured S3/S0 data (purple). e) Snapshot of the induced surface charge distribution at different times, where the arrow indicates the rotation of the surface charge in the *xz* plane over time.

The observed π/2 phase difference between the degenerate transversal dipolar modes $p_x$ and $p_z$ is in full agreement with the model described above, with the driving electric fields of the electron in quadrature (Eqn. (1)). The coherent excitation of $p_x$ and $p_z$ with a π/2 phase delay then creates a rotating dipole in the *xz* plane where the light spin angular momentum (SAM) $\mathbf{S}_p = \mathrm{Im}(\mathbf{p}^* \times \mathbf{p})$ points in the y direction. This phase delay is clear in Figure 3c that shows the time evolution of the induced dipole moments upon electron excitation; the delay is constant over the full plasmon relaxation time. Similarly, in Figure 3d it is clear that the phase difference between $p_x$ and $p_z$ remains constant in the time domain. To further illustrate the origin of the observed SAM we show a snapshot of the induced surface charge distribution in the time domain in Figure 3e. The snapshot indicates the unidirectional charge rotation in the *x-z* plane with a direction that is set by the phase of the dipole moments excited by the electron beam. The inversion of the angular polarization distribution then directly follows from the $\sigma_{yz}$ related excitation position, leading to a counter-propagating surface charge distribution, which was used to calculate the induced dipole moments. Our data show that interference between resonant modes of different order (e.g. dipolar and quadrupolar, as in work by Chi et al.[28] ) is not a necessary condition to generate CPL from a single plasmonic nanoantenna.

## CPL generation by controlling transversal- and longitudinal modal interference

Next, we study the generation of CPL as a result of coherently exciting transversal- and longitudinal dipolar modes, by exciting the nanorod in an aloof configuration with the electron beam at a distance of 5 nm from the nanorod tip. The spectral overlap between the resonances (see Figure 1b) enables interference between the coherently excited transversal and longitudinal dipolar modes. Figure 4 shows the experimental angle-resolved CL polarimetry data, together with simulations at a wavelength of 650±25nm. The same experiment with the electron beam positioned at the other tip is presented in the SI (Figure S9), and shows mirror-imaged angle-resolved CL intensities, as expected.

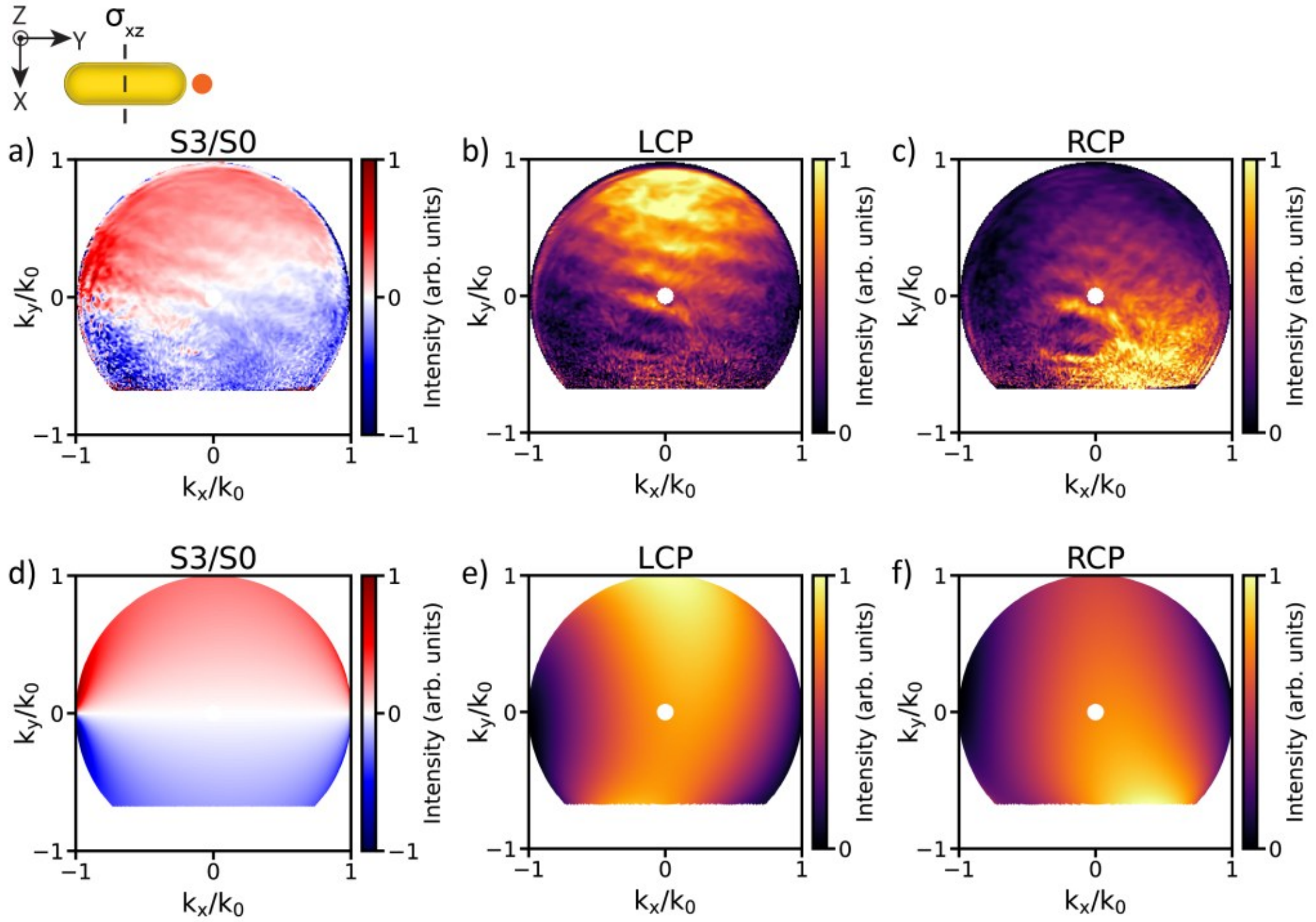

Figure 4. Experimental (a-c) and simulated (d-f) angle-resolved CL data for 30 keV aloof electron excitation of the Au nanorod at the tip (orange, 5 nm from the surface), using a bandpass filter at 650±25nm. (a,d) Normalized S3 Stokes parameter, (b,e) left-handed circularly polarized (LCP) and (c,f) right-handed circularly polarized (RCP) CL intensity.

For this geometry the optical spin-Hall effect is also observed, now with the LCP and RCP components showing distinct sideways beaming with a symmetry axis along the longitudinal axis of the particle, perpendicular to the direction observed in Figure 3 for excitation at the side of the nanoparticle. The experiments and simulations are in good agreement, aside from a small rotation of the experimental angle-resolved CL images that we ascribe to the nanorod not being perfectly perpendicular to the optical axis. We find that scaling in the simulations the amplitude of $p_z$ by a factor of 2.53 and shifting the phase by -0.28π is required to reproduce the experimental results. As above we attribute these discrepancies to a slight spatial misalignments of the electron beam, and the presence of the substrate and the creation of a thin carbon layer at the particle surface. The full angle-resolved polarimetry dataset and fit results can be found in the SI (section 5)

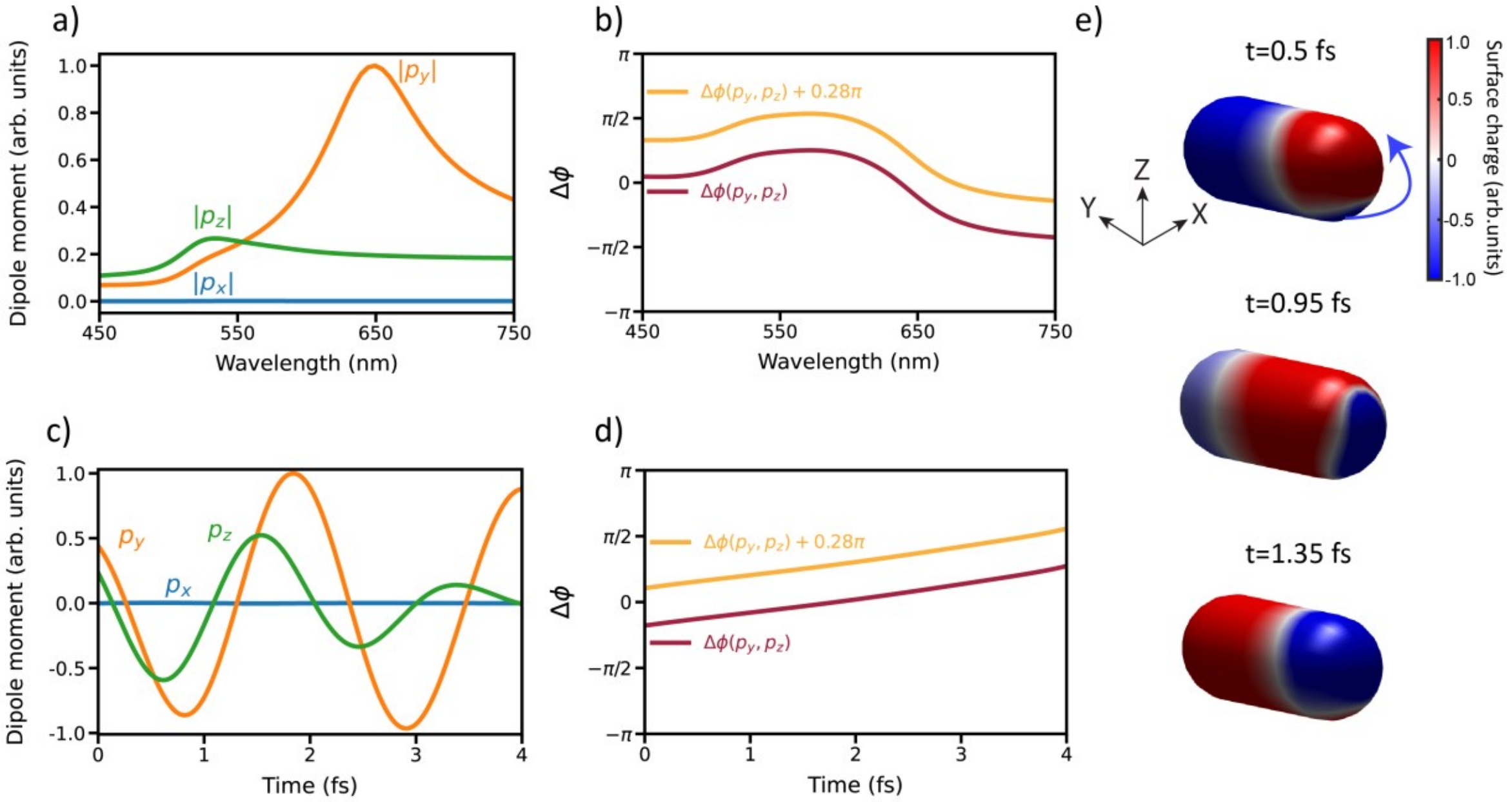

Figure 5. a) BEM-calculated dipole moment spectra for the three dipole orientations for 30 keV aloof electron excitation of the Au nanorod at the tip (orange dot in Figure 4, 5 nm from the surface). Data for $p_z$ are scaled to match the measured S3/S0 data. b) Spectra of phase difference between $p_y$ and $p_z$, as calculated (red) and corrected to match the measured S3/S0 data (yellow). c) Time evolution of normalized $p_x$, $p_y$, $p_z$ with t=0 defined as the point in time where the electron is closest to the nanostructure. d) Time evolution of the phase difference between $p_y$ and $p_z$ as calculated with BEM (red) and corrected to match the measured S3/S0 data (yellow). e) Snapshot of the induced surface charge distribution at different times, where the arrow indicates the rotation of the surface charge in the *yz* plane over time.

To illustrate the origin of the OSHE for this geometry we show the calculated induced dipole moments and their phase differences in the frequency and time domain in Figure 5. For this geometry the transverse mode $p_x$ is not exited as it does not couple to the radial and transverse electron beam fields ($p_x$=0). We observe a strong spectral contribution of both the longitudinal $p_y$ mode at 650 nm and the transverse $p_z$ mode at 535 nm. Figure 5b shows the calculated phase difference between $p_z$ and $p_y$ as well as the shifted phase spectrum found by fitting the experimental angular profile for S3/S0. The phase shifts vary significantly over the frequency range where the CL is collected (650±25nm), contrary to the results discussed in the previous section where the π/2 relationship was nearly constant over the entire frequency range. This is due to the fact that we excite two non-degenerate modes, each with a different phase variation with frequency. This phase delay is clear in Figure 5c which shows the time evolution of the induced dipole moments upon electron excitation; Figure 5d shows that the phase gradually increases with increasing time, which is a consequence of the more intricate dependence of the phase difference with wavelength as shown in Figure 5b, combined with the fact that the modes have different relaxation times. To further illustrate the origin of the observed OSHE we show a snapshot of the induced surface charge distribution in the time domain in Figure 5e; and illustrate the full unidirectional charge rotation in the *yz* plane. The antisymmetric angular polarization distributions then directly follow from the antisymmetric excitation positions, leading to counter-propagating surface charge distribution corresponding to the calculated induced dipole moments.

## Creating chiral CL by symmetry breaking

Finally, we present experiments performed by exciting the nanorod at impact parameters that do not coincide with the $\sigma_{xz}$ and $\sigma_{yz}$ mirror planes. In Figure 6a-d we show the angle-resolved S3/S0 images for four excitation positions indicated by the colored dots in the schematic in the center of the figure.

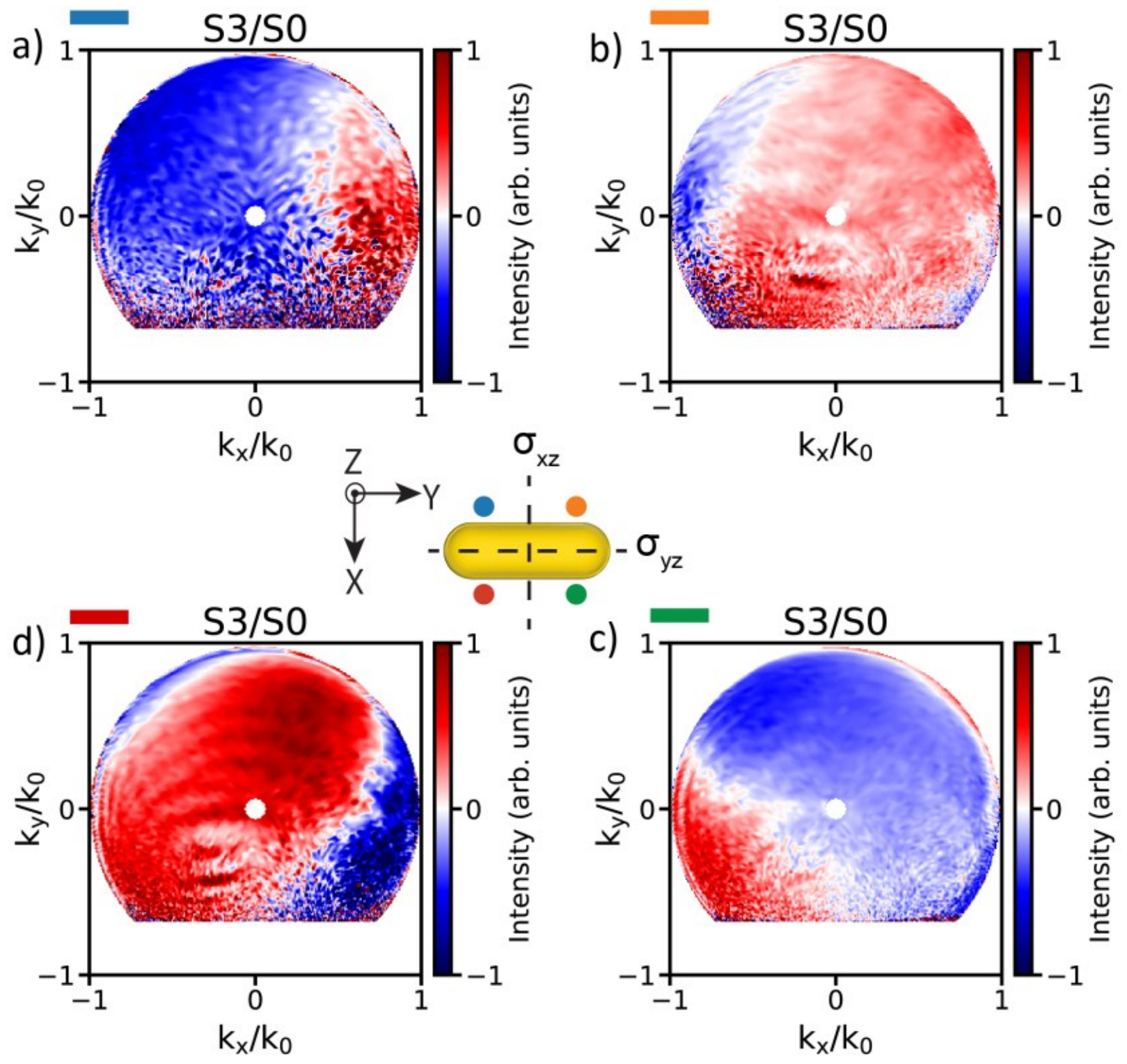

Figure 6. Experimental angle-resolved S3/S0 images (650±25 nm) measured by exciting the nanorod with 30 keV electrons at four positions off the symmetry planes (5 nm from the surface). The colored lines above the panels correspond to the impact parameter indicated by the colored dots in the schematic in the center.

Figure 6 shows that the CL has a preferred handedness in the collected angles for all beam locations, as shown by the predominance of either LCP or RCP in the collected emission angles. We express the chiral nature of the collected CL emission by the integrated DOCP, defined as $DOCP_{int} = \frac{\sum_i S3_i}{\sum_i S0_i}$ , where *i* represents each pixel in the image. We calculate the integrated $DOCP_{int}$ to be -0.24 ,0.15, -0.14 and 0.33 for the results in Figure 6a-d, respectively, with a negative sign corresponding to preferential emission of RCP and a positive sign LCP preferentially emitted. We find sign changes and mirror symmetries in the angular emission profiles around by the $\sigma_{xz}$ and $\sigma_{yz}$ mirror planes, in a similar fashion as discussed above Angle-resolved LCP and RCP images that correspond to the data shown in Figure 6 are shown in the SI (section 6), along with the calculated results.

In Figure 7 we present the calculated induced dipole moments and the relative phase differences associated with the excitation position indicated by the red dot in Figure 6. Exciting the Au nanorod away from the symmetry planes results in the excitation of dipole moments in all three directions $p_x$, $p_y$ and $p_z$ (Figure 7a). Distinct phase differences are observed between the induced dipole components (Figure 7b) which we again attribute to the intricate dependence of the phase difference on wavelength, combined with the fact that the modes have different relaxation times, with the $p_y$ component having the longest relaxation time (Figure 7c). As in Figure 5 this also leads to variations in the relative phase with time (Figure 7d). The induced surface charge distributions shown in Figure 7e reflect the complex excitation of the combined dipole moments with the surface charges rotating over both the *xz* and *xy* plane.

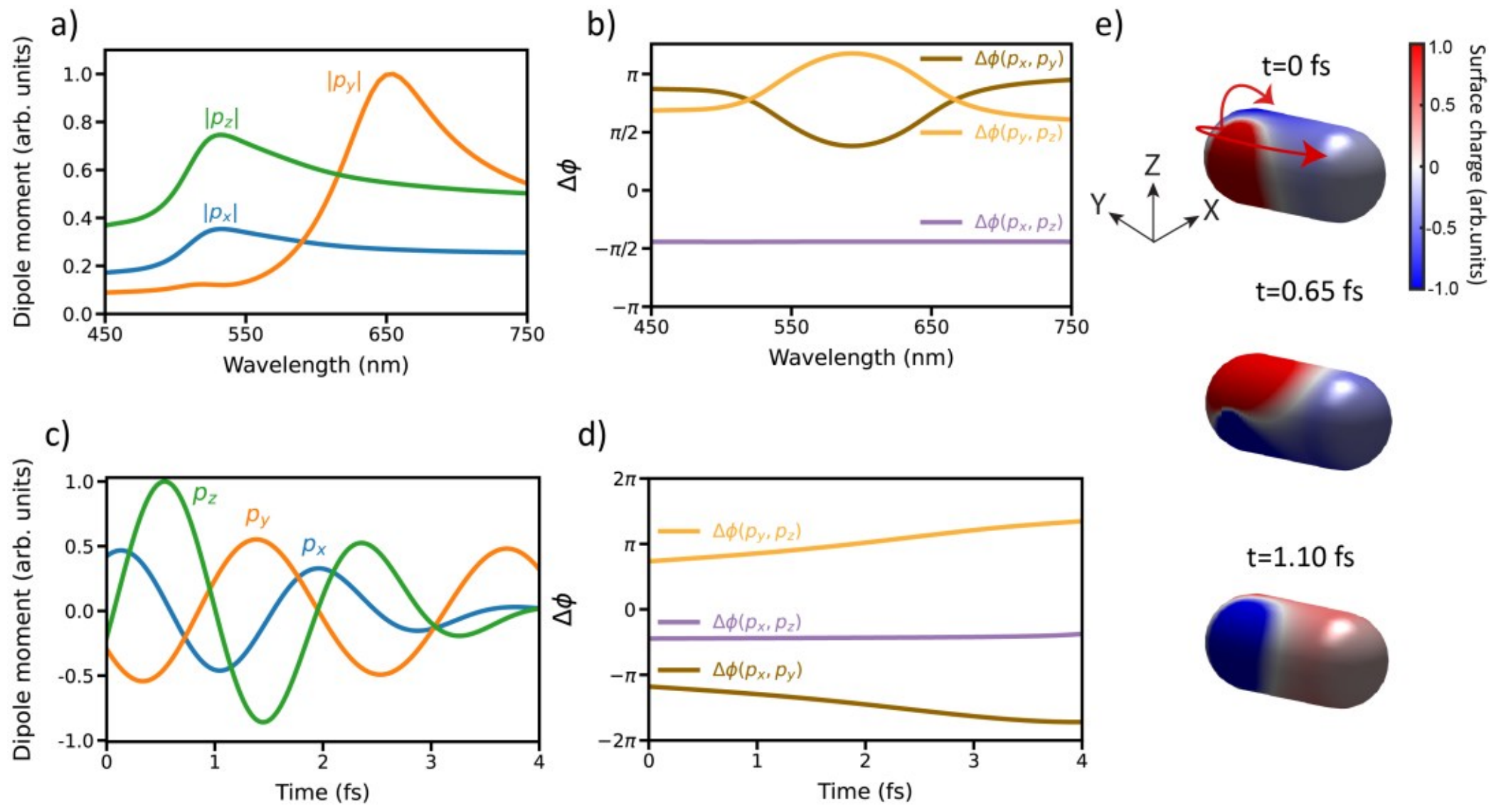

Figure 7. a) BEM-calculated induced dipole moment spectra for the three dipole orientations for 30 keV electron excitation at the position indicated by the red dot in Figure 6 (5 nm from the surface). Data are scaled to match the measured S3/S0 data. b) Spectra of phase differences between the three dipole components, corrected to match the measured S3/S0 data. (c) Time evolution of normalized $p_x$, $p_y$, $p_z$ with t=0 defined as the point in time where the electron is closest to the nanostructure. d) Time evolution of the phase difference between the three dipole components calculated with BEM and corrected to match the measured S3/S0 data. e) Snapshots of the induced surface charged distributions at different times, where the red arrows indicate the rotation of the surface charge around the nanoparticle.

## Conclusion

We showed how 30-keV electrons can be used to control the generation of CPL and chiral CL from a single plasmonic nanoantenna. Taking advantage of the $\pi/2$ phase difference between the transverse and longitudinal components of the electric field that is carried by the electron, coherently excited dipolar modes interfere in the far field and create an optical spin-Hall effect. The angular distribution of the emitted CPL and the degree of chirality are controlled by the position of the electron beam relative to the nanoparticle. Our findings represent an elemental form of generation of CPL, by synthesizing the circularly rotating electric fields characteristic for CPL from interference of orthogonal dipolar modes with controlled phase lag.

Several potential applications come to mind. Using specially designed nanoparticle arrays, polarization-controlled and chiral Smith-Purcell emission may be achieved that can find applications in advanced electron spectrophotometry. Applications in chiral sensing will require specially tailored sample sensing geometries in the electron microscope. We note the concepts demonstrated here are also applicable for low energy electrons (~100 eV) for which alternative practical and simpler geometries can be designed. For potential applications in communication, displays and quantum communication, this work may inspire novel concepts in which precisely located (single-)photon sources are placed at selective locations on nanostructures, thereby creating chiral (single-photon) emission. Furthermore, the presented femtosecond time domain snapshots of controlled charge distribution across the metal surface may inspire novel concepts for controlled femto-catalysis, where subsequent reactions pathways are controlled at time scales within an optical cycle.

# Methods

## Synthesis

To obtain Au nanorods, small Au seeds were synthesized by adding 5 mL of 0.5 mM $HAuCl_4$ and 5 mL of 0.2 M CTAB in a round bottom flask. Subsequently, 600 µL of ice cold 0.01M $NaBH_4$ and 400 µL milliQ (MQ) water were added to the flask while rigorously stirring the solution. The seed solution was then aged at room temperature for 30 minutes, and stored for later use.

1.125 g of CTAB and 191.75 mg of NaOL were dissolved in 31.25 mL of 50°C MQ water in a round bottom flask, after which the solution allowed to cool down to 30°C. Subsequently, 1.5 mL of 4 mM $AgNO_3$ was added, and the solution was left undisturbed for 15 minutes. Afterwards, 31.25 mL of 1 mM $HAuCl_4$ was added and the solution was left for 90 minutes until it became colorless. In the next step 140 µL of concentrated HCl was added to adjust the pH of the solution to 1.51 and the mixture was stirred at 400 rpm. Next, 156 µL of 64 mM ascorbic acid solution was added and the mixture was stirred for 30 seconds. Finally, 1.25 µL of the earlier made seed solution was added to the flask, and the solution was stirred for another 30 seconds. The solution was left undisturbed at 30°C for 12 hours for the growth of the nanorods.

## Sample preparation

The as-synthesized nanorod solution was centrifuged at 5000 rpm for 5 minutes after which the supernatant was removed, this step was repeated 3 times. The solution was then diluted by a factor of 5, after which 20 µL was dropcasted onto a 15 nm $Si_3N_4$ TEM grid (TED pella), which was made hydrophobic beforehand, by a 10 minute UV ozone cleaning step. The grid was then transferred to an $O_2$ plasma cleaner (Diener Electronic), where the sample was exposed to a 100W $O_2$ plasma for 15 minutes.

# CL measurements

## Nanoparticles

All measurements were performed on the nanorods synthesized as described in the synthesis section. Due to gradual degradation of the CL signal as a result of carbon deposition under electron beam exposure, it was not feasible to perform all measurements on the same nanoparticle. Therefore, the measurements were performed on a set of 8 nanoparticles with approximately the same size, aspect ratio, and orientation with respect to the CL collection mirror.

## CL spectroscopy

All CL measurements were performed in an Apreo2C SEM (Thermo Fisher Scientific inc., MA, USA) with a Schottky electron source. The CL collection was done on a SPARC system (DELMIC BV, The Netherlands), where a parabolic mirror was placed between the sample and the electron pole piece and subsequently the light was directed to a spectrometer. The acceleration voltage was set at 30 keV and the beam current at 1.6 nA with an acquisition time of 0.5 s per pixel. The CL spectra were corrected for the background CL emission from the $SI_3N_4$ substrate, and the collection efficiency of the detector.

The angle-resolved CL polarimetry measurements were performed with an acceleration voltage of 30 keV and beam current of 1.6 nA. An aluminum (Al) parabolic mirror collects the emitted light, and directs it towards a polarizer equipped with a quarter wave plate (QWP) and a linear polarizer (LP), after which the light is projected onto a 2D CCD detector. The geometry of the parabolic mirror allows the light emission angle to be retrieved from the image projected on the CCD, as each emission angle corresponds to a unique pixel on the CCD detector.

The acquisition time was 15-45 s per polarization filtered channel, leading to a total acquisition time of 90-270 s. The choice of exposure time depended on the CL intensity of the nanorod and the intensity of the resonance spectra. Measurements with an acquisition time of 15 s. per polarization channel were performed in experiments where the longitudinal mode was observed. Measurements involving the excitation of only the transversal dipolar modes were done with an acquisition time of 45 s. per polarization filtered channel due to relatively weak CL emission. The pixel size was selected to be between 4-7 nm, and the optical bandpass filters used in this work were 550±25nm and 650±25nm.

The measured polarization-filtered images are projected onto a spherical coordinate system where components $E_\theta$ and $E_\phi$ are the transverse basis, and correspond to p- and s- polarized light emission in this coordinate system, respectively. For comparing experimental results with simulations we adopt the same coordinate system, where the calculated electric far field in cartesian space is projected onto the spherical basis defined by (θ,ϕ).

We apply the inverse Mueller matrix to the calculated results to directly compare the experimental angle-resolved CL polarimetry results with the calculated results. Angle-resolved CL polarimetry results for experiments and simulations are presented in the same basis. Stokes parameters are presented in the sample plane basis, which is the relevant representation basis for polarization information, in contrast to the polarization filtered angle-resolved intensity distributions, which are presented as collected on the detector.

**Definfitions of the Stokes parameters**

We define the Stokes parameters from the observer perspective where one looks from the camera towards the parabolic mirror, which is adopted throughout this work. The horizontal (H) and vertical (V) polarization components correspond the y- and z oriented electric field components respectively.

$$S0 = I_H + I_V = |E_y|^2 + |E_z|^2$$

$$S1 = I_H - I_V = |E_y|^2 - |E_z|^2$$

$$S2 = I_{45} - I_{135} = 2\mathrm{Re}\left(E_y E_z^*\right)$$

$$S3 = I_{LHC} - I_{RHC} = -2\mathrm{Im}\left(E_y E_z^*\right) \quad (7)$$

## Simulations

All simulations were performed using the BEM implementation in the MNPBEM17 MATLAB toolbox. The Au nanorods were 50x110 nm, had a total of 1084 triangular face elements, and the optical constants used for gold were taken from Johnson et al.[36] The spectrum was simulated from 1.5 to 3 eV (826 nm to 413 nm) to reproduce our experimental frequency range of the optical system. For all our simulations we set the effective refractive index of the environment to n=1.4, in order to account for the presence of the $Si_3N_4$ substrate and a thin hydrocarbon layer at the particle surface, which alter the dielectric environment of the nanorod. The electron energy was set at 30 keV, the lateral beam width was set at 0.1 nm, and the impact parameter was defined at a distance of 5 nm from the surface at each respective excitation position. From the results, the induced surface charges, cathodoluminescence spectra, and angle-resolved CL polarimetry results were calculated. The time domain results presented in this work were calculated by performing the Inverse fast Fourier transform (IFFT) on the calculated frequency domain data.

## Author contributions

S.P. fabricated the samples, performed the CL measurements, the BEM simulations, and did the data analysis. H.M assisted with the data analysis and reviewed the manuscript. Q.N. synthesized the Au nanorods and recorded the TEM image. A.R.E contributed to the data analysis. S.P., H.M and A.P. wrote the original manuscript, Q.N and A.R.E coedited the manuscript. All authors provided feedback and contributed to the manuscript.

## Notes

The authors declare the following competing financial interest(s): Albert Polman is cofounder and co-owner of Delmic BV, a company that produces commercial cathodoluminescence systems like the one that was used in this work.

## Acknowledgments

This project has received funding from the European Research Council (ERC) under the European Union's Grant Agreement No. 101019932 (Quantum Electron Wavepacket Spectroscopy, QEWS) and Grant Agreement No. 101017720 (EBEAM).

# Supplementary Information

# Generating Circularly Polarized and Chiral Cathodoluminescence from a Single Plasmonic Nanoantenna

Stan de Peinder*, Hollie Marks, Quynh Nguyen, Álvaro Rodríguez Echarri and Albert Polman

Center for Nanophotonics, NWO-Institute AMOLF, Science Park 104, 1098XG Amsterdam, The Netherlands

*Corresponding author: s.dpeinder@amolf.nl

Contents

# 1 Electric dipole modes

Figure S1 shows the surface charge distribution of the electric dipolar modes present in a gold nanorod. Here, $p_x$ and $p_z$ indicate the transversal dipolar modes with a resonance at 535 nm, shown as the blue spectrum in Figure 1b in the main text, and the longitudinal dipolar mode, $p_y$, peaking at 650 nm, shown in orange. The surface charge distributions are calculated by exciting the nanorod with a planewave at the respective resonance frequency, polarized along the direction of the displayed dipole moment.

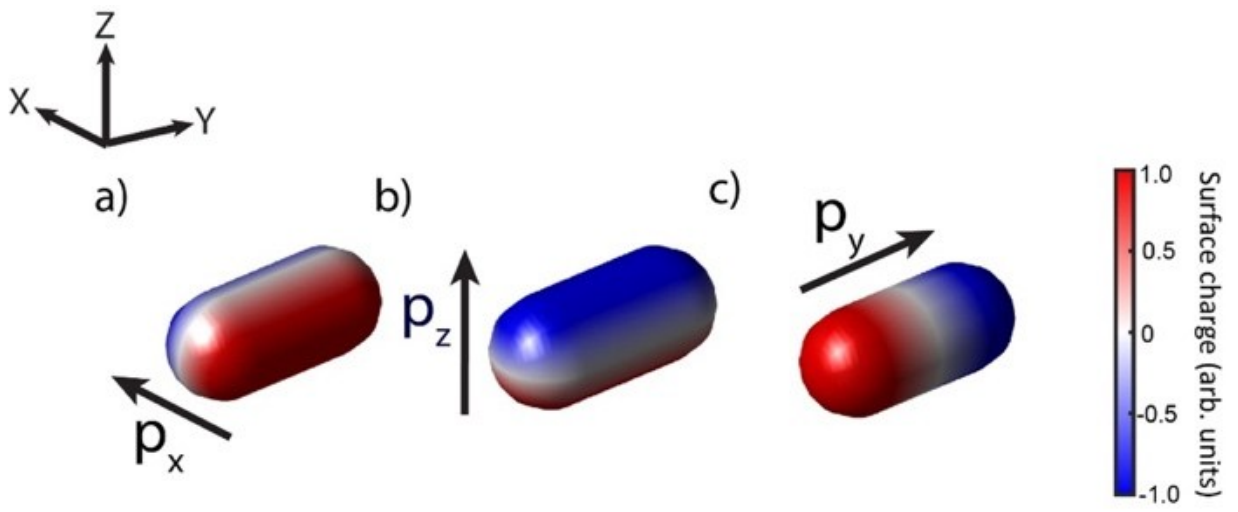

Figure S1. BEM simulations of electric dipole modes in a gold nanorod. (a,b) Surface charge distributions of transversal dipole modes defined as $p_x$ and $p_z$. (c) Surface charge distribution of the longitudinal dipole mode defined as $p_y$.

# 2 Mirror-symmetric excitation of transversal dipole modes

Figure S2 shows the angle-resolved CL polarimetry results for the experiment where the nanorod is excited at the $\sigma_{yz}$ related impact parameter, compared to the results shown in Figure 2 in the main text. We show the symmetric nature of the CPL generation in the far field, which is mirror-imaged due to excitation at the opposite side of the nanorod.

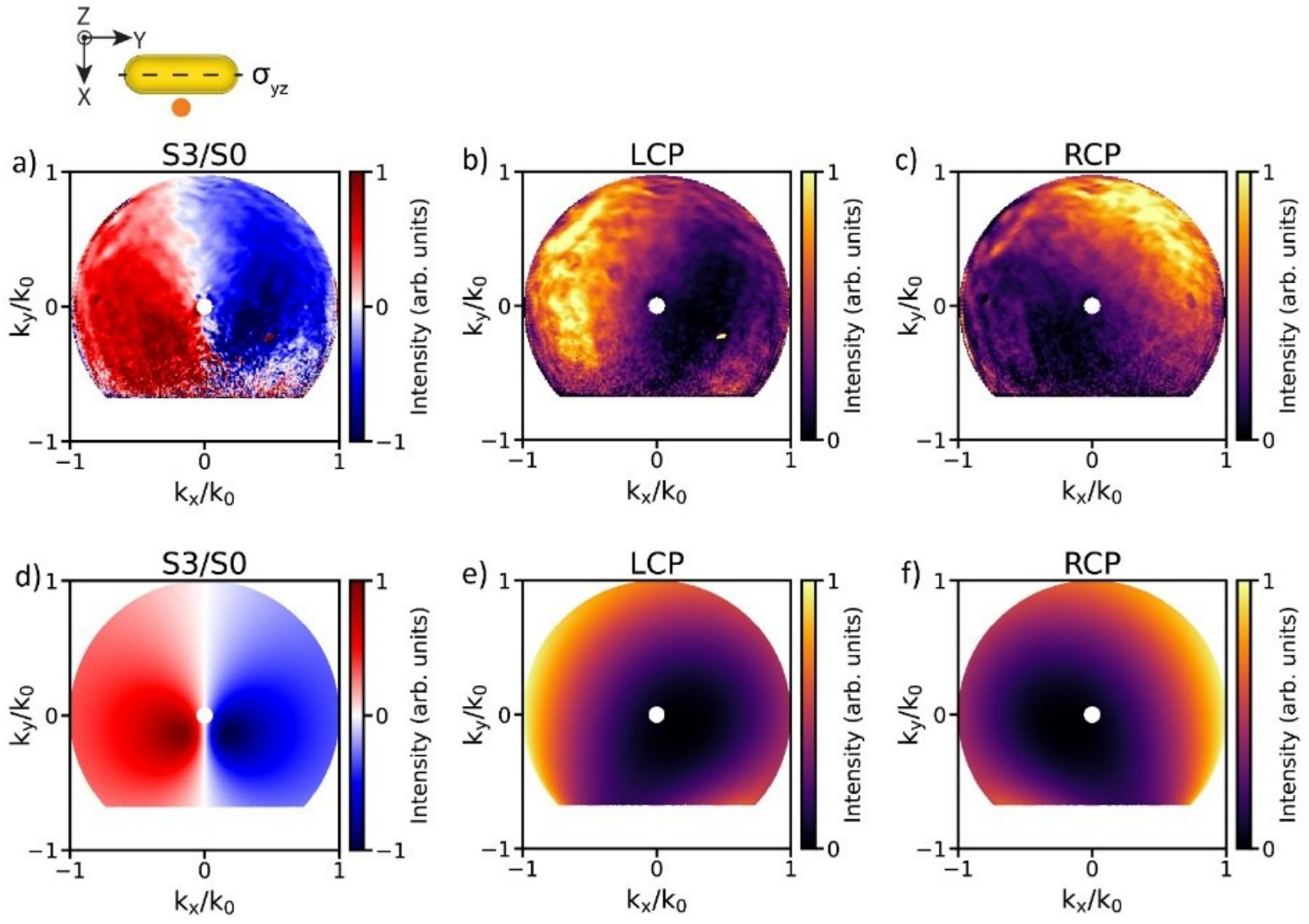

Figure S2. Experimental and calculated angle-resolved CL data for 30 keV aloof electron excitation of the Au nanorod at the indicated positions (orange dot), using a bandpass filter at 550±25nm. (a,d) Normalized S3 Stokes parameter, (b,e) left-handed circularly polarized (LCP) and (c,f) right-handed circularly polarized (RCP) CL intensity.

# 3 Angle-resolved CL polarimetry results for interfering transversal dipole modes

## 3.1 Experimental angle-resolved CL polarimetry

Figure S3 shows the polarization- and angle-resolved CL results measured in the experiment presented in Figure 2 in the main text. We show all measured data here; we use all polarization- and angle-resolved images for fitting our BEM calculations to the experimental results.

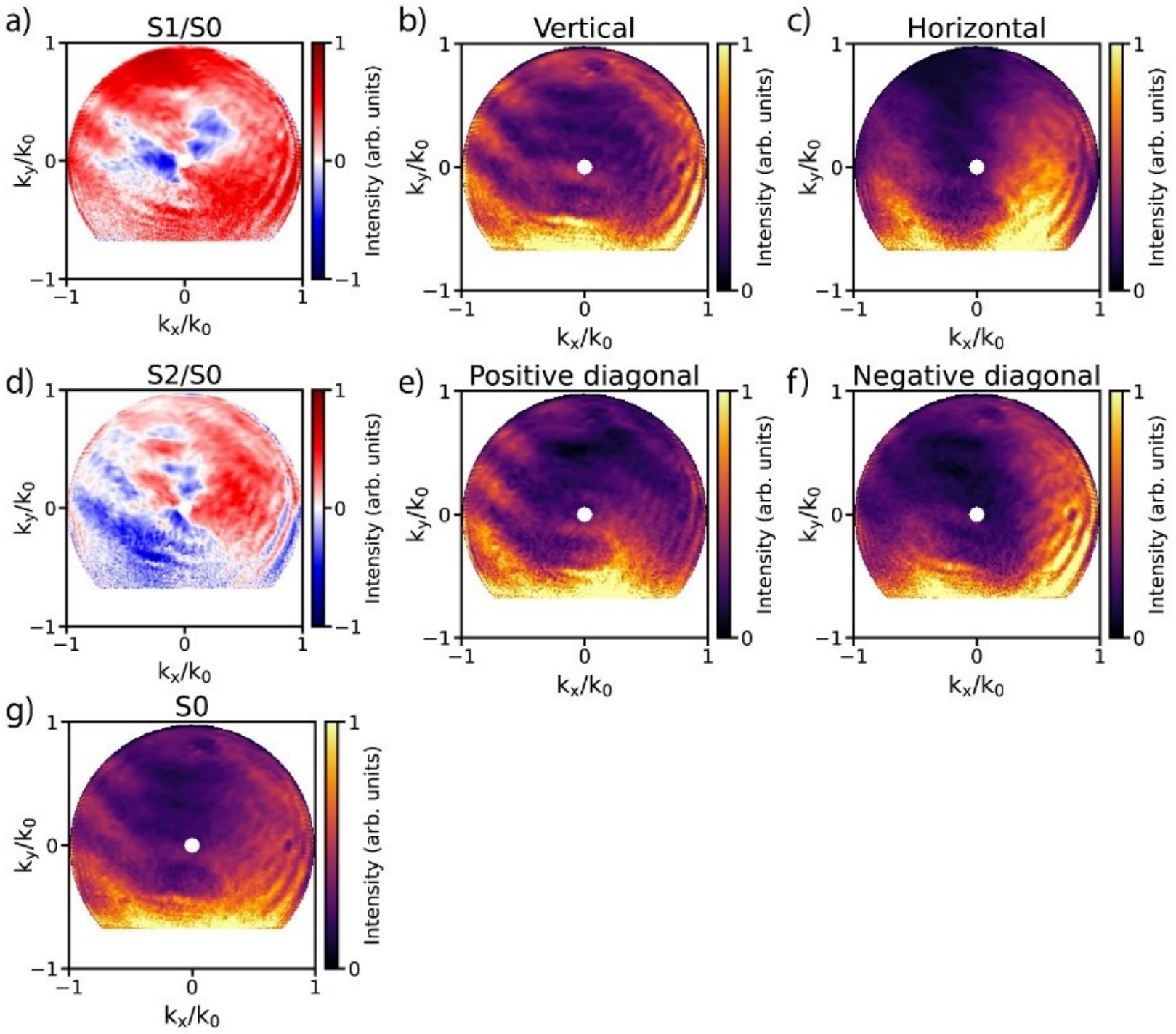

Figure S3. Experimental angle-resolved polarization-resolved CL emission (a) Normalized Stokes parameter $S_1/S_0$. (b,c) Angle-resolved linear vertically and horizontally polarized CL emission. d) Normalized Stokes parameter $S_2/S_0$. (e,f) Angle-resolved positive-diagonal (+45°) and negative-diagonal (−45°) polarized CL emission. (g) $S_0$ parameter, used to normalize $S_1$ to $S_3$.

## 3.2 Calculated angle-resolved CL polarimetry

Figure S4 shows the calculated polarization- and angle-resolved CL intensity results using our dipole approximation model. Subsequently, we fit the calculated complex dipole moments to our experimental data. The fit procedure is discussed in section 3.3, where we also show the fit results in Table 1.

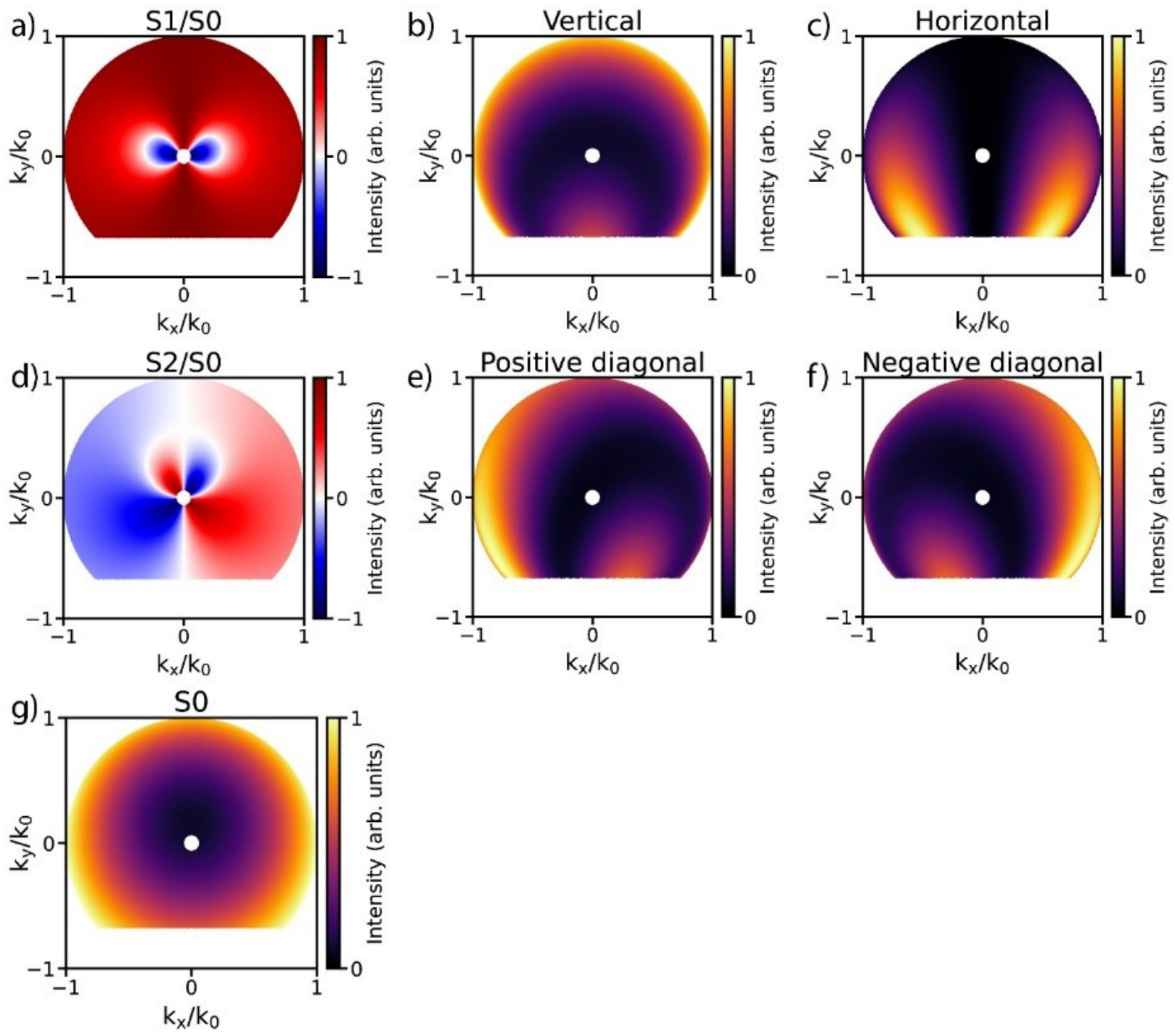

Figure S4. BEM-calculated angle-resolved polarization-resolved CL emission**.** (a) Normalized Stokes parameter S1/S0 (b,c) Angle-resolved vertically and horizontally polarized CL emission. (d) Normalized Stokes parameter S2/S0. (e,f) Angle-resolved positive-diagonal (+45°) and negative-diagonal (−45°) polarized CL emission. (g) S0 parameter, used to normalize S1 to S3.

## 3.3 Stokes parameter fit procedure

To fit our calculated angle-resolved CL polarimetry results to the experiment results, we use the *fminsearch* function in MATLAB. In the case where we find that two dipole components contribute to our far field, we fix the amplitude and phase of one component. For the second component, the amplitude is allowed to change to values [0.2, 10] with intervals of 0.02, allowing us to retrieve an optimal amplitude ratio. The phase is allowed to range from [-π,π] with intervals of 0.01π. The root mean square error (RMSE) between the experimental and simulated normalized Stokes parameters is then calculated at every step, until the improvement of the calculated error is <0.1%.

$$\mathrm{RMSE}_{S_i/S_0} = \sqrt{\frac{\sum_{n\in\mathcal{M}_i} w_n \left[\left(\frac{S_i}{S_0}\right)_n^{\mathrm{sim}} - \left(\frac{S_i}{S_0}\right)_n^{\mathrm{exp}}\right]^2}{\sum_{n\in\mathcal{M}_i} w_n}}, \qquad i \in \{1,2,3\}. \qquad \text{(Eq. S1)}$$

From our fit procedure we retrieve values for the amplitude ratio scaling factors ($c_z/c_x$), and phase shift correction factors defined as, ($\Delta\phi$ ($c_z$-$c_x$)/π), and report the $c_z$ interval which describes the dipole amplitude scaling values that produce fit results within 1% of the optimum value shown in the $c_z/c_x$ cell. Hereby, we show that the effective scaling that is applied to reproduce our experiments, can take a range of values. We choose to perform our fits on the normalized Stokes parameters, as the fit procedure yielded better results compared to performing the fit on the raw polarization filtered angle-resolved images. We attribute the reduction in fit quality using the raw detected photon counts to the reduction in the intensity of CL emission during the acquisition. Small carbon deposition, as a result of electron beam exposure, results in a reduction in CL emission over the course of the measurement. Because the polarization- and angle-resolved images are measured in succession, the CL photon count rate reduces over the course of the measurements, producing an additional error in intensity that reduces the fit quality. We find that performing the fit on the normalized Stokes parameter reduces this error. Finally, we report the calculated RMSE, to describe the discrepancy between our calculations and experiments. We attribute the discrepancies to differences in the amplitudes, the presence of interferences in experiment which are not accounted for in our calculations, and slight misalignments of the nanoparticle with respect to the optical axis of the system. In this way, we collect values for the parameters as summarized for the data in Figure 4 in Table 1 .

| **$\|c_z/c_x\|$** | **$\Delta\phi$ ($c_z$-$c_x$)** | **$c_z$ interval** |
|---|---|---|
| 3.94 | 0.15 π | [3.60, 4.40] |
| **S1/S0 RMSE** | **S2/S0 RMSE** | **S3/S0 RMSE** |
| 0.43 | 0.32 | 0.32 |

Table 1. Fit results for the simulations shown in Figure S4. Top row (left-to-right), relative amplitude weighting constants used for $p_x/p_z$, the calculated relative phase shift correction $\phi$($c_z$-$c_x$), and the interval in which the amplitude weighting reproduces a fit within 1% of the found optimal value. Bottom row (left to right), RMSE calculated from fitting our calculated S1/S0, S2/S0, and S3/S0 to the experimental results shown in Figure S3.

# 4 Angle-resolved CL polarimetry results for mirror symmetric excitation of transversal dipole modes.

## 4.1 Experimental angle-resolved CL polarimetry

Figure S5 shows the polarization- and angle-resolved CL data measured for the experimental results presented in Figure S2.

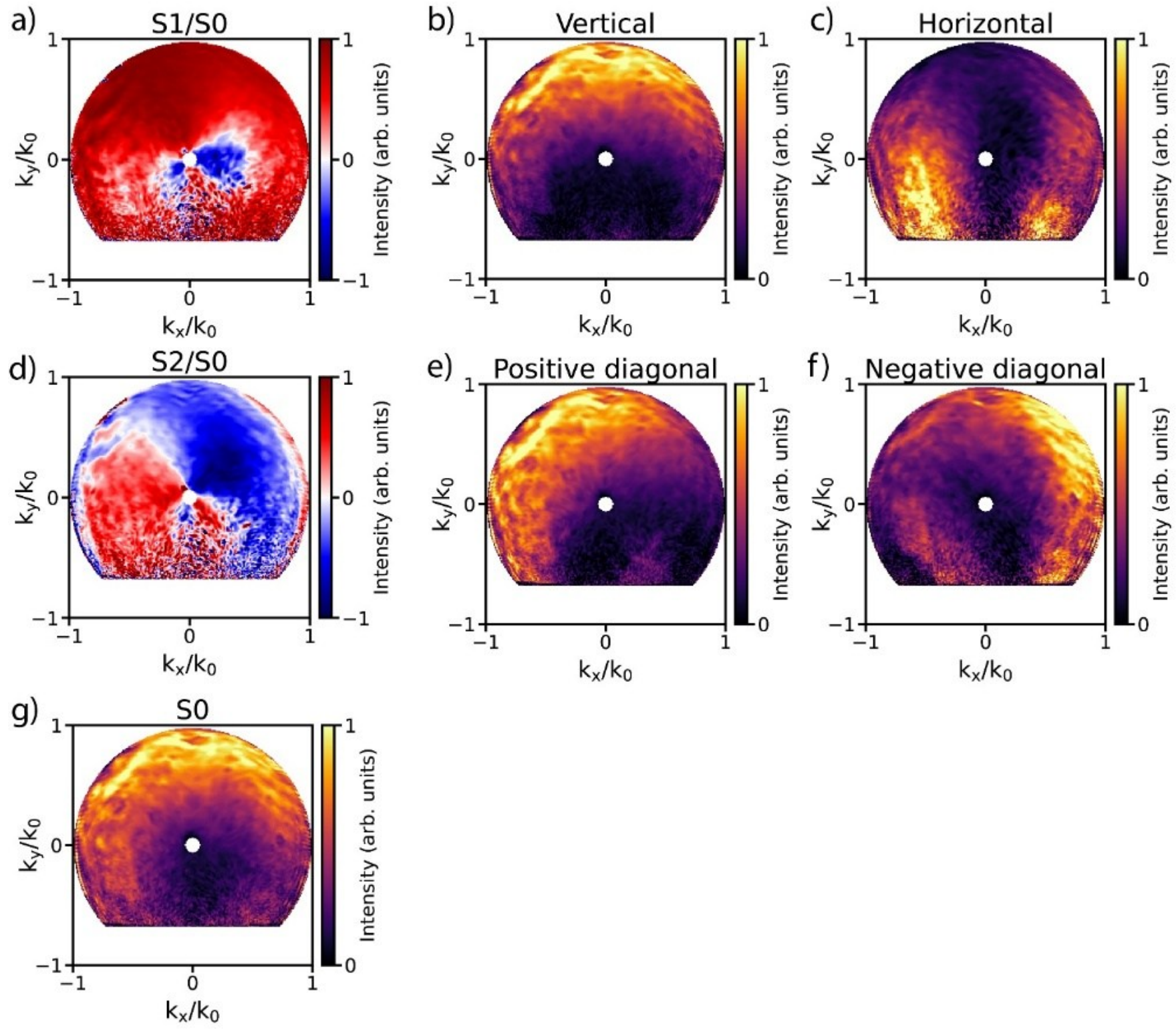

Figure S5. Experimental angle-resolved polarization-resolved CL emission (a) Normalized Stokes parameter S1/S0. (b,c) Angle-resolved linear vertically and horizontally polarized CL emission. d) Normalized Stokes parameter S2/S0. (e,f) Angle-resolved positive-diagonal (+45°) and negative-diagonal (−45°) polarized CL emission. (g) S0 parameter, used to normalize S1 to S3.

## 4.2 Calculated angle-resolved CL polarimetry

In Figure S6 we show calculated polarization- and angle-resolved CL emission results to support our experimental results shown in Figure S2. We use our dipole approximation model and subsequently fit the calculated complex dipole moments to our experimental data. The fit results are shown in Table 2.

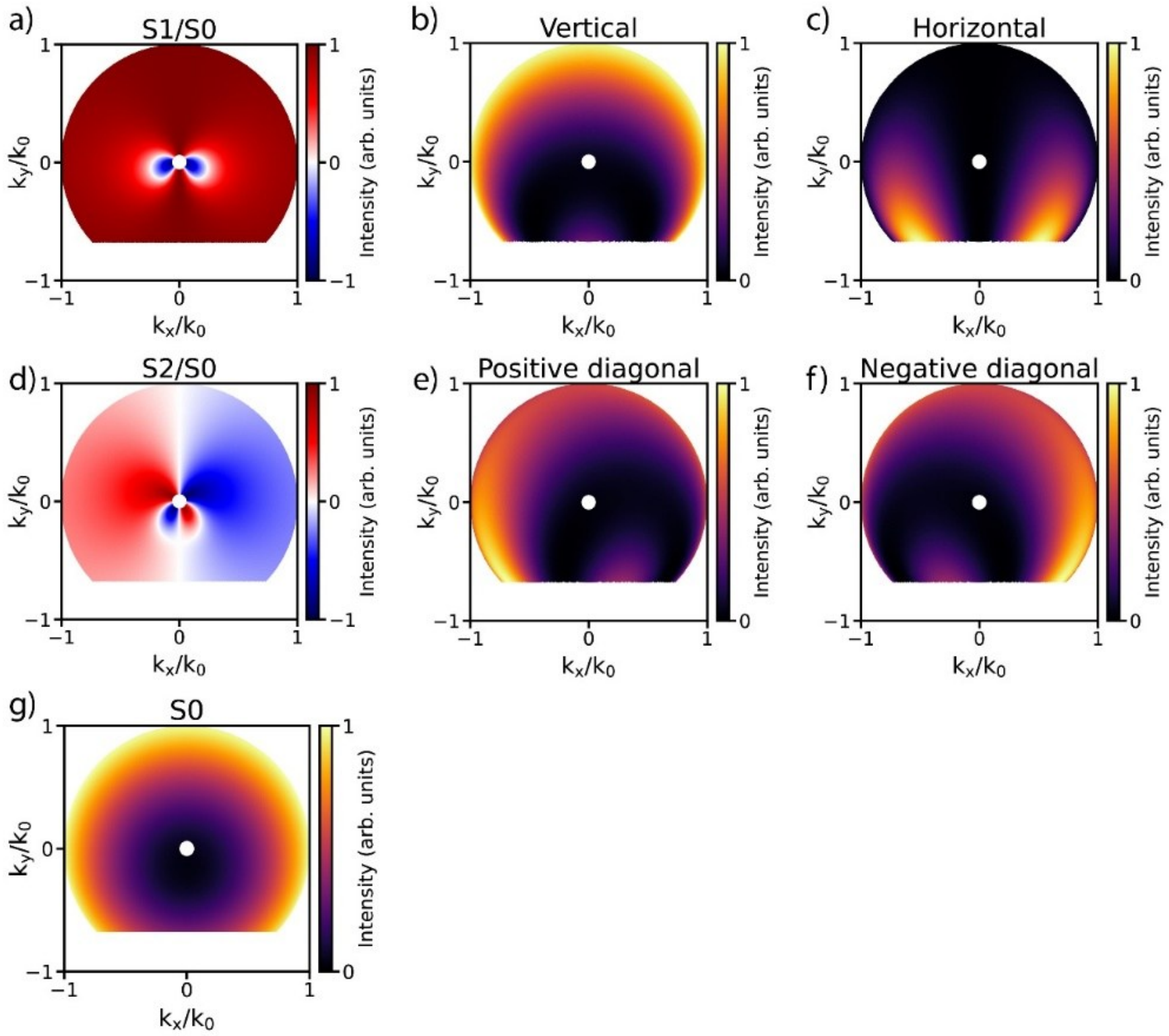

Figure S6. BEM-calculated angle-resolved polarization-resolved CL emission**.** (a) Normalized Stokes parameter S1/S0 (b,c) Angle-resolved vertically and horizontally polarized CL emission. (d) Normalized Stokes parameter S2/S0. (e,f) Angle-resolved positive-diagonal (+45°) and negative-diagonal (−45°) polarized CL emission. (g) S0 parameter, used to normalize S1 to S3.

| **$\|C_z/C_x\|$** | **Δϕ ($c_z$-$c_x$)** | **$C_z$ interval** |
|---|---|---|
| 5.54 | 0.20 π | [5.20, 6.00] |
| **S1/S0 RMSE** | **S2/S0 RMSE** | **S3/S0 RMSE** |
| 0.26 | 0.23 | 0.20 |

Table 2. Fit results for the calculated results shown in Figure S6. Top row (left to right), optimal relative amplitude weighting ratio of $p_z/p_x$, the calculated relative phase shift correction ϕ($c_z$-$c_x$), and interval in which the amplitude weighting reproduces a fit within 1% of the found optimal value. Bottom row: RMSE calculated from fitting our calculated S1/S0, S2/S0, and S3/S0 to the experimental results shown in Figure S5.

# 5 Angle-resolved CL polarimetry results for interfering transversal and longitudinal dipole modes.

## 5.1 Experimental angle-resolved CL polarimetry

In Figure 4 we show the measured circular polarization channels and the normalized S3 parameter when the nanorod is excited at the nanorod tip. In Figure S7, we show all other polarization- and angle-resolved CL intensity images which are used for our fitting procedure.

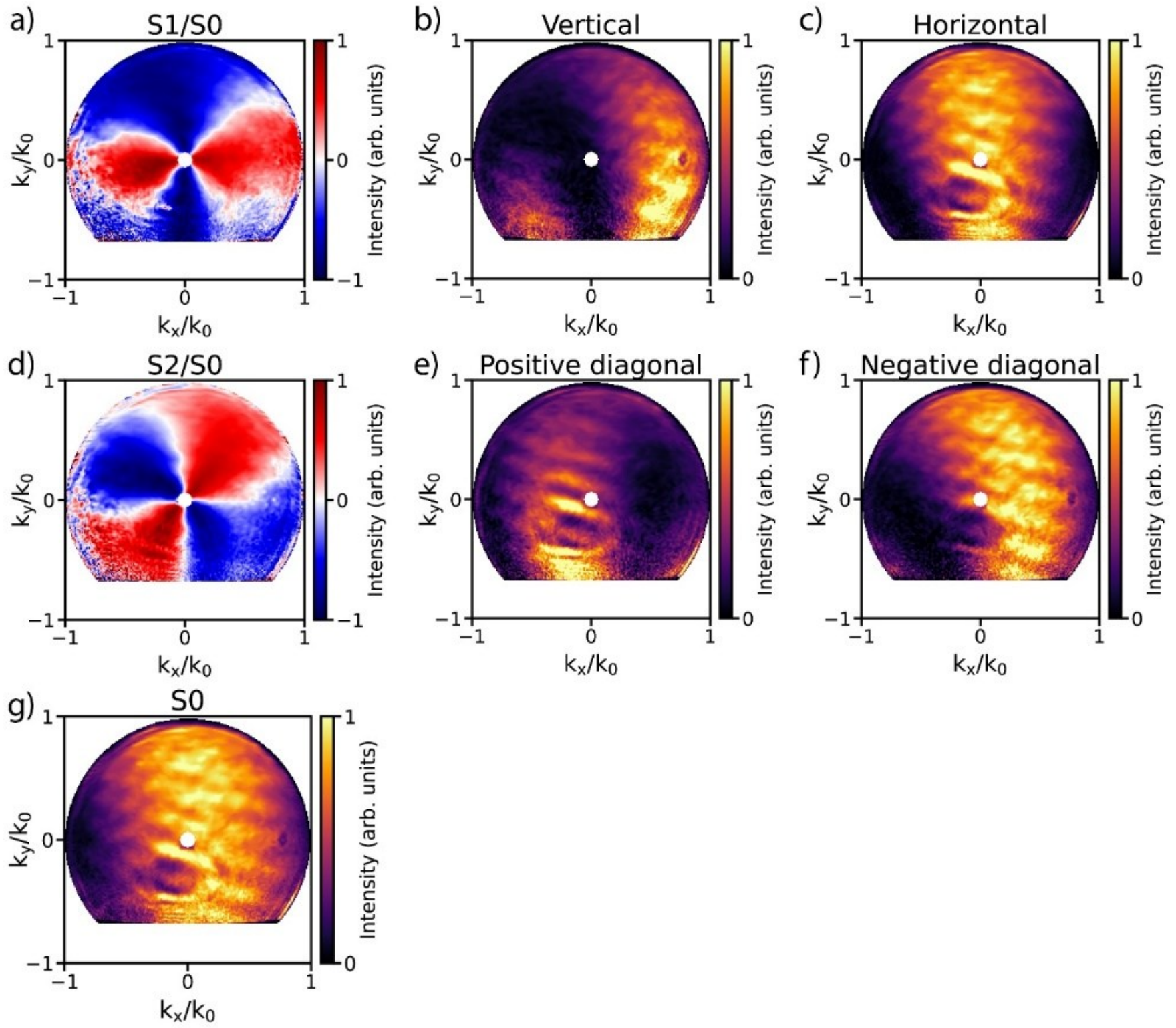

Figure S7. Experimental angle-resolved polarization-resolved CL emission (a) Normalized Stokes parameter S1/S0. (b,c) Angle-resolved linear vertically and horizontally polarized CL emission. d) Normalized Stokes parameter S2/S0. (e,f) Angle-resolved positive-diagonal (+45°) and negative-diagonal (−45°) polarized CL emission. (g) S0 parameter, used to normalize S1 to S3.

## 5.2 Calculated angle-resolved CL polarimetry

In Figure 4 we show the angle-resolved S3/S0 parameter, and the angle-resolved LCP and RCP intensity distributions measured when the nanorod was excited at the nanorod tip. In Figure S8, we show the calculated results, and the fit results are shown in Table 3.

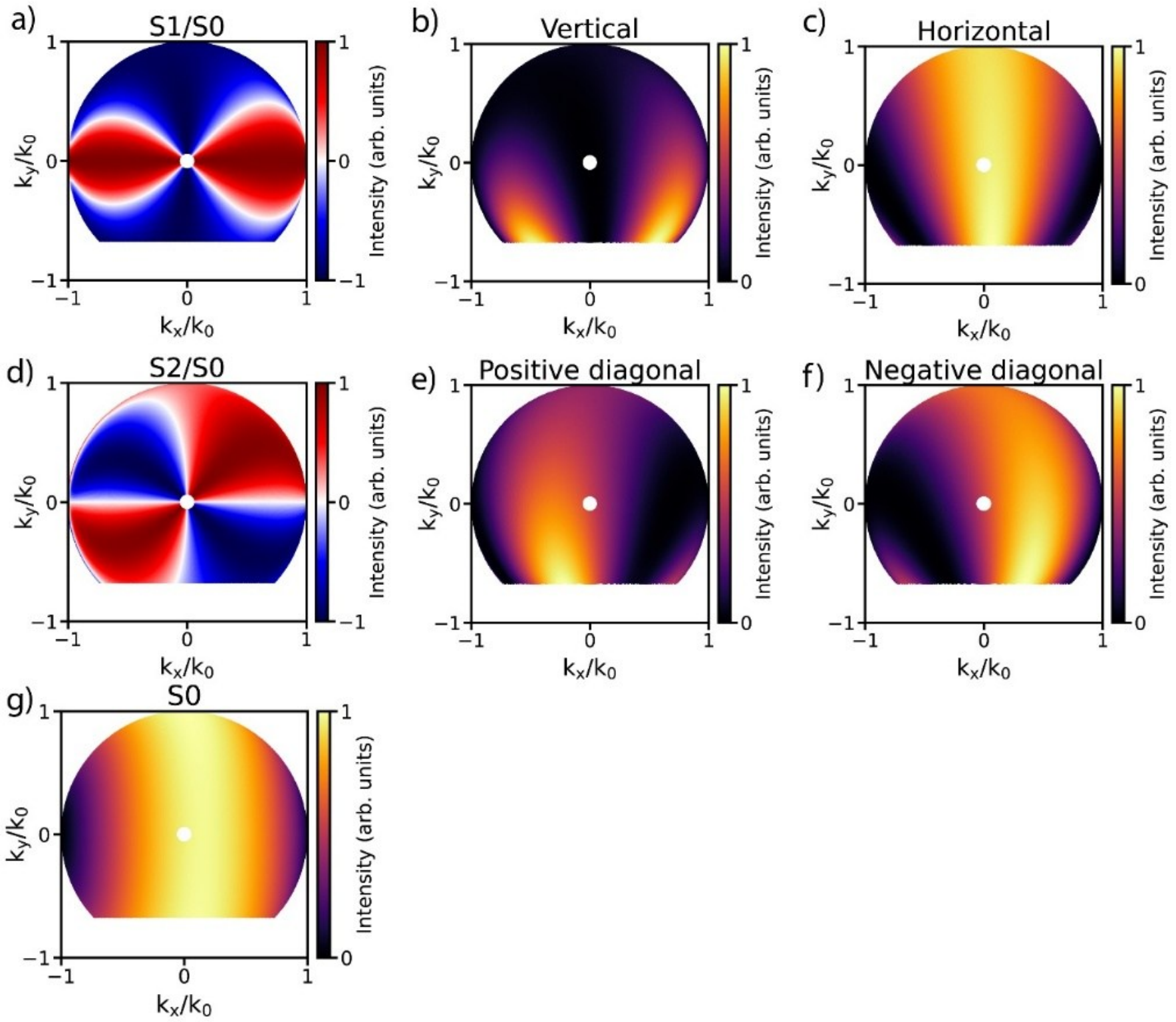

Figure S8. BEM-calculated angle-resolved polarization-resolved CL emission**.** (a) Normalized Stokes parameter S1/S0 (b,c) Angle-resolved vertically and horizontally polarized CL emission. (d) Normalized Stokes parameter S2/S0. (e,f) Angle-resolved positive-diagonal (+45°) and negative-diagonal (−45°) polarized CL emission. (g) S0 parameter, used to normalize S1 to S3.

| **$\|C_z/C_y\|$** | **Δϕ ($c_z$-$c_y$)** | **$C_z$ interval** |
|---|---|---|
| 2.54 | -0.28 π | [2.23, 2.85] |
| **S1/S0 RMSE** | **S2/S0 RMSE** | **S3/S0 RMSE** |
| 0.28 | 0.31 | 0.10 |

Table 3. Fit results for the calculated results shown in Figure S8. Top row (left-to-right), optimal relative amplitude weighting ratio of $p_z/p_y$, the calculated relative phase shift correction ϕ($c_z$-$c_y$), and interval in which the amplitude weighting reproduces a fit within 1% of the found optimal value. Bottom row: RMSE calculated from the best fit for fitting calculated S1/S0, S2/S0, and S3/S0 to the experimental results.

## 5.3 Mirror-symmetric excitation of interfering transversal and longitudinal dipole modes

Figure S9 shows the results for the experiment and calculations where the nanorod is excited at the $\sigma_{xz}$ related symmetry impact parameter, compared to Figure 5 in the main text. We note that there are discrepancies between experiment and simulation, especially in the S3/S0 image, which we attribute to slight misalignment of the nanorod with respect to the optical axis, and poor signal-to-noise ratio. Nevertheless, the main trends regarding the mirror-symmetry-related changes are clearly visible in the angle-resolved LCP and RCP images.

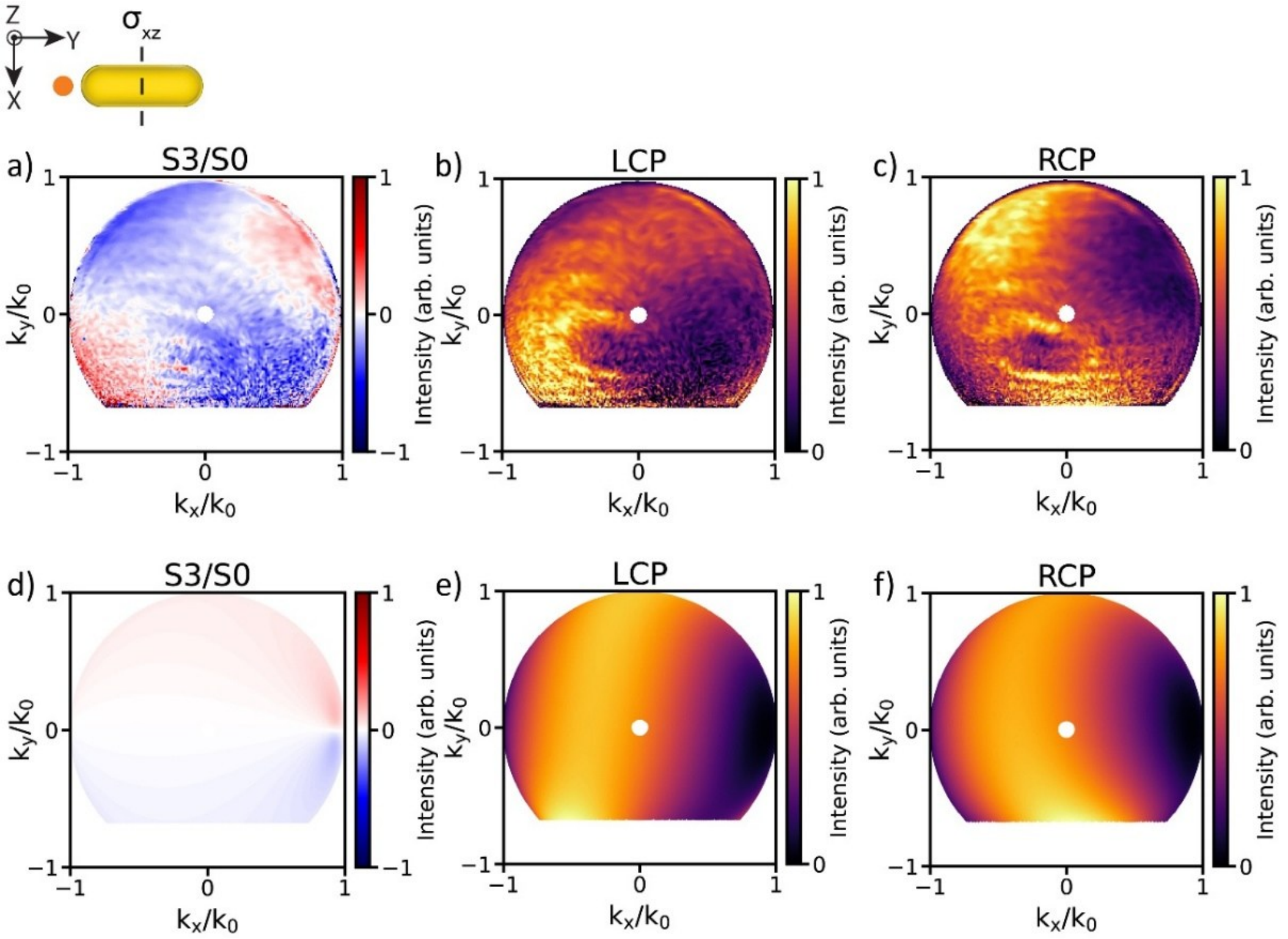

Figure S9. a) Normalized S1 parameter. b) linear vertical polarized CL emission. c) Horizontal polarized CL emission. d) Normalized S2 parameter. e) Positive diagonal polarized CL. f) negative diagonal polarized light emission. g) S0 parameter used to normalized the angle-resolved Stokes parameters.

| **$\lvert C_z/C_y \rvert$** | **$\Delta\phi$ ($c_z$-$c_y$)** | **$C_z$ interval** |
|---|---|---|
| 5.00 | -0.04 π | [4.45, 5.00] |
| **S1/S0 RMSE** | **S2/S0 RMSE** | **S3/S0 RMSE** |
| 0.25 | 0.27 | 0.16 |

Table 4. Fit results for the calculated results shown in Figure S9. Top row (left-to-right), optimal relative amplitude weighting ratio of $p_z/p_y$, the calculated relative phase shift correction $\phi(c_z$-$c_y)$, and interval in which the amplitude weighting reproduces a fit within 1% of the found optimal value. Bottom row: RMSE calculated from the best fit for fitting calculated S1/S0, S2/S0, and S3/S0 to the experimental results.

## 6. Angle-resolved CL polarimetry results for chiral CL generation

In Figure 6 in the main text we show the generation of chiral CL at four different excitation positions. In this section we show the polarization- and angle-resolved CL emission intensity distributions that were measured for each respective dataset. We show the experimental results in Figure S10, and the fit results in Table 5.

### 6.1.1 Experimental angle-resolved CL polarimetry (Figure 6a)

In this section we show the experimental- and calculated angle-resolved CL polarimetry results that are part of the data collected for the results shown in Figure 6a.

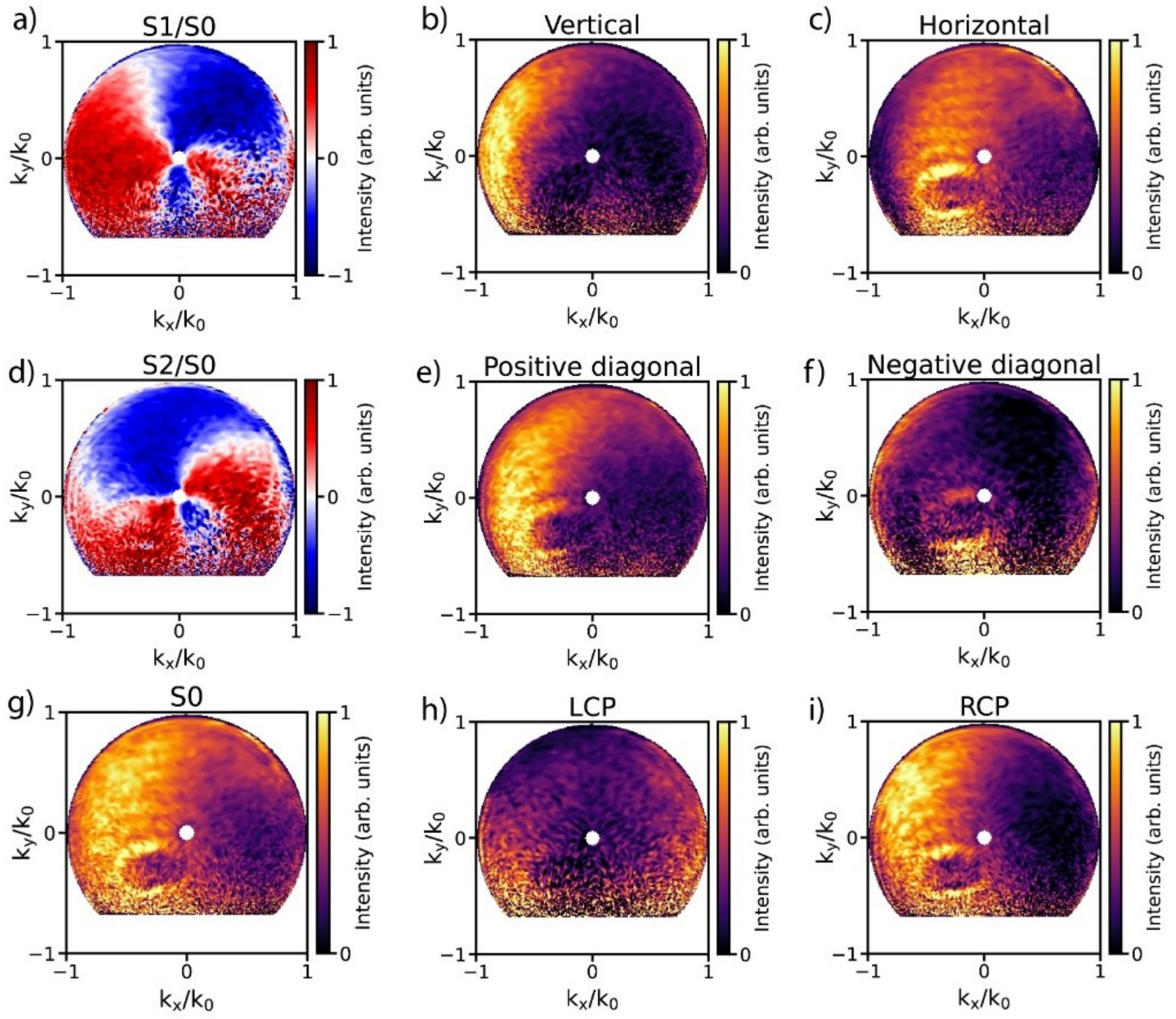

Figure S10. Experimental angle-resolved polarization-resolved CL emission (a) Normalized Stokes parameter S1/S0. (b,c) Angle-resolved linear vertically and horizontally polarized CL emission. d) Normalized Stokes parameter S2/S0. (e,f) Angle-resolved positive-diagonal (+45°) and negative-diagonal (−45°) polarized CL emission. (g) S0 parameter, used to normalize S1 to S3. (h,i) Angle-resolved left- and right handed circularly polarized CL emission.

## 6.1.2 Calculated angle-resolved CL polarimetry (Figure 6a)

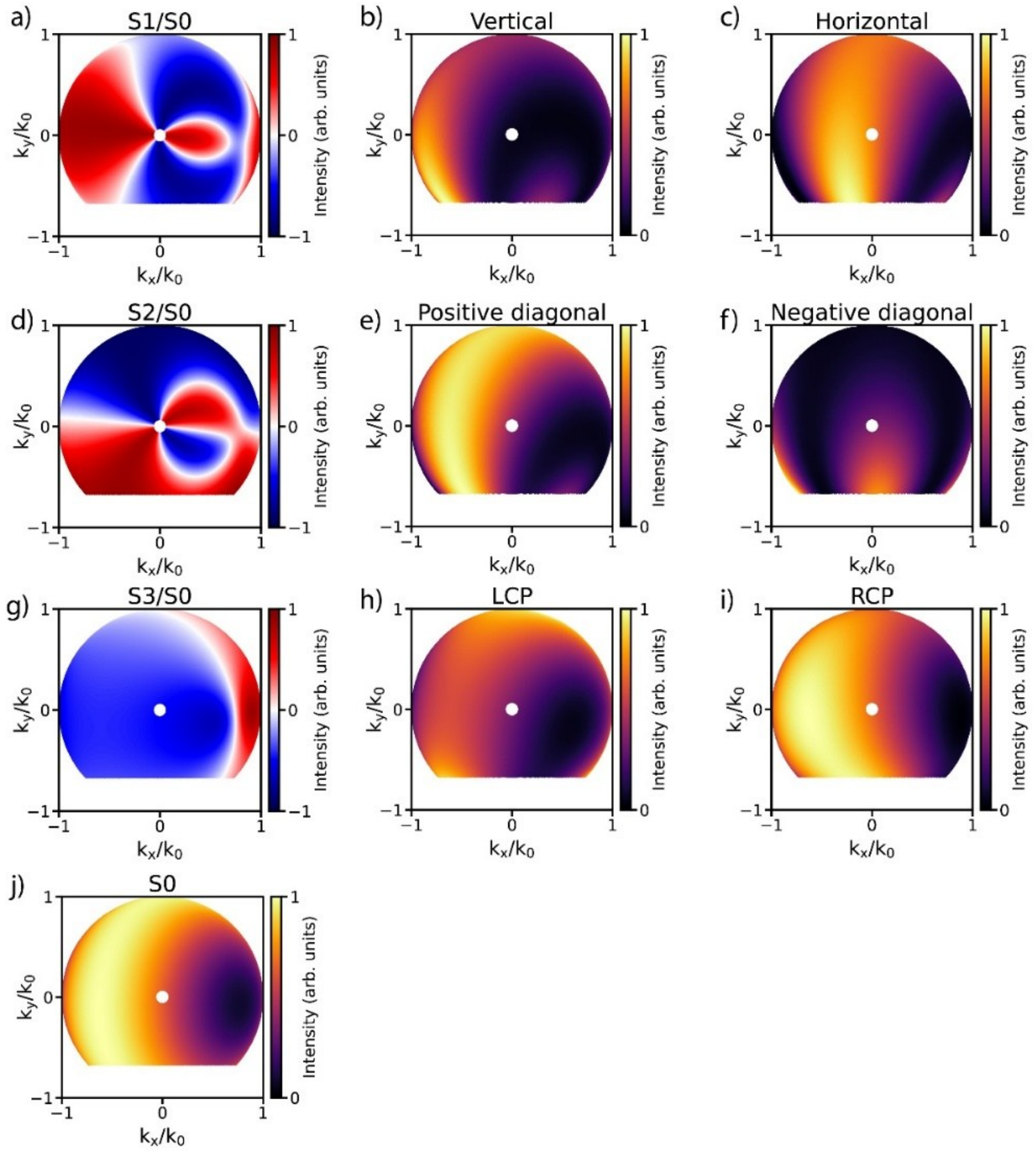

Figure S11. BEM-calculated angle-resolved polarization-resolved CL emission**.** (a) Normalized Stokes parameter S1/S0 (b,c) Angle-resolved vertically and horizontally polarized CL emission. (d) Normalized Stokes parameter S2/S0. (e,f) Angle-resolved positive-diagonal (+45°) and negative-diagonal (−45°) polarized CL emission. (g) Normalized Stokes parameter S3/S0. (h,i) Left- and right handed circularly polarized CL emission. (j) Stokes parameter S0, used for normalization.

| **$\|C_z/C_y\|$** | **Δϕ ($c_z$-$c_y$)** | **$C_z$ interval** |
|---|---|---|
| 3.66 | -0.08 π | [3.20, 4.00] |
| **S1/S0 RMSE** | **S2/S0 RMSE** | **S3/S0 RMSE** |
| 0.35 | 0.41 | 0.31 |

Table 5. Fit results for the calculated results shown in Figure S11. Top row (left-to-right), optimal relative amplitude weighting ratio of $p_z/p_y$, the calculated relative phase shift correction ϕ($c_z$-$c_y$), and interval in which the amplitude weighting reproduces a fit within 1% of the found optimal value. Bottom row: RMSE calculated from fitting our calculated S1/S0, S2/S0, and S3/S0 to the experimental results.

### 6.2.1 Experimental angle-resolved CL polarimetry (Figure 6b)

In this section we show the experimental- and calculated angle-resolved CL polarimetry results that are part of the data collected for the results shown in Figure 6b.

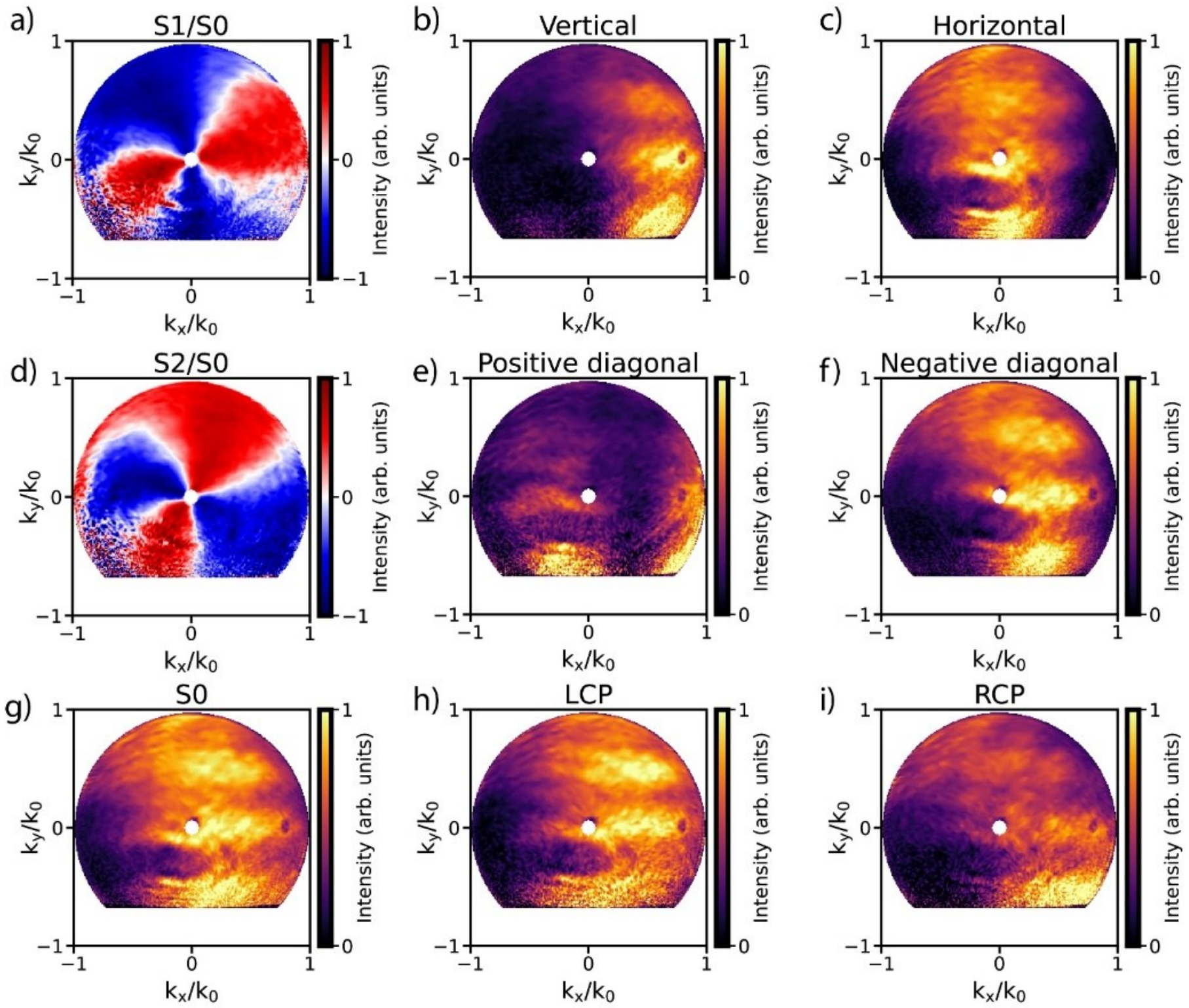

Figure S12. Experimental angle-resolved polarization-resolved CL emission (a) Normalized Stokes parameter S1/S0. (b,c) Angle-resolved linear vertically and horizontally polarized CL emission. d) Normalized Stokes parameter S2/S0. (e,f) Angle-resolved positive-diagonal (+45°) and negative-diagonal (−45°) polarized CL emission. (g) S0 parameter, used to normalize S1 to S3. (h,i) Angle-resolved left- and right handed circularly polarized CL emission.

## 6.2.2 Calculated angle-resolved CL polarimetry (Figure 6b)

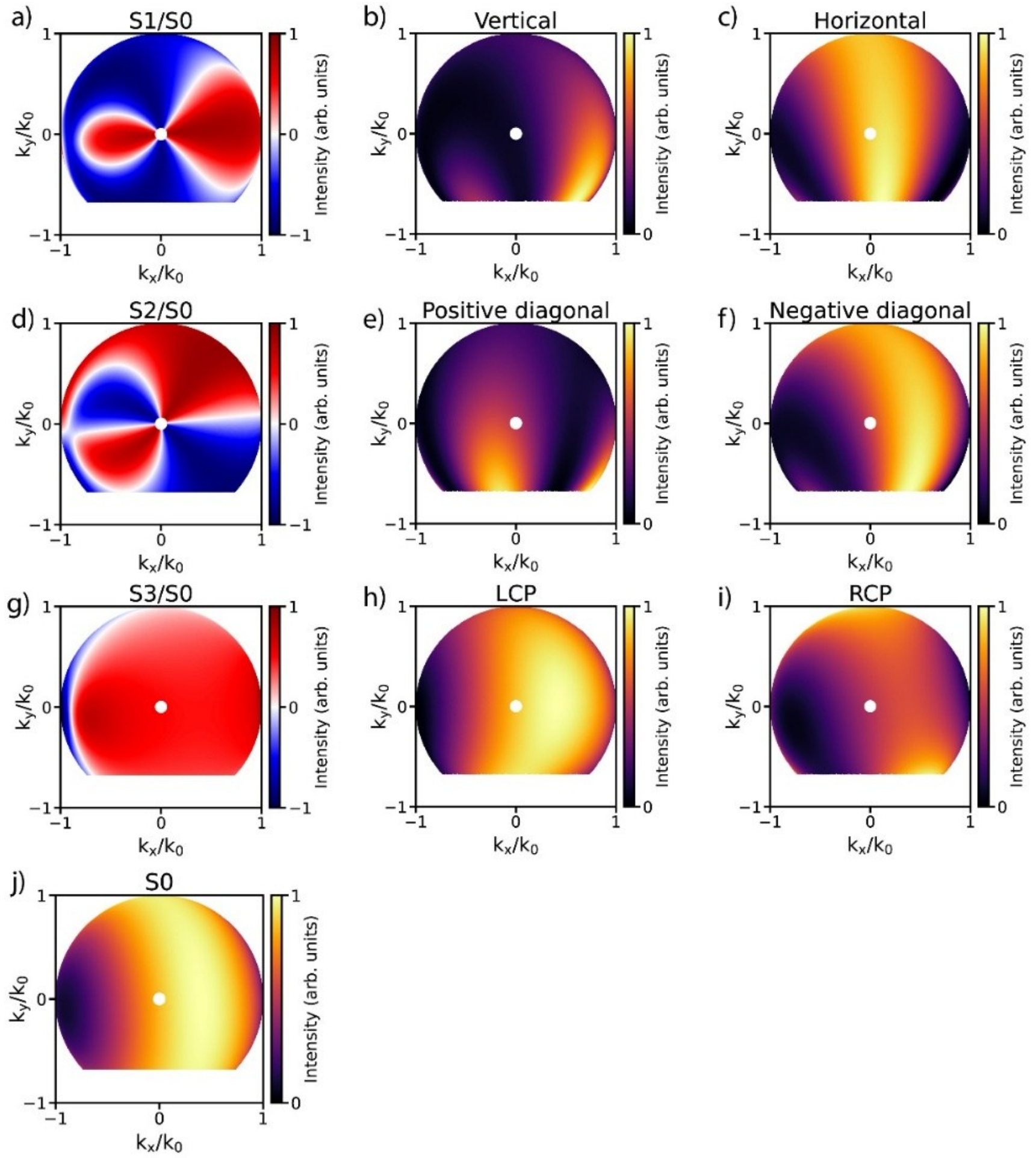

Figure S13. BEM-calculated angle-resolved polarization-resolved CL emission**.** (a) Normalized Stokes parameter S1/S0 (b,c) Angle-resolved vertically and horizontally polarized CL emission. (d) Normalized Stokes parameter S2/S0. (e,f) Angle-resolved positive-diagonal (+45°) and negative-diagonal (−45°) polarized CL emission. (g) Normalized Stokes parameter S3/S0. (h,i) Left- and right handed circularly polarized CL emission. (j) Stokes parameter S0, used for normalization

| **$\lvert C_z/C_y \rvert$** | **$\Delta\phi$ ($c_z$-$c_y$)** | **$C_z$ interval** |
|---|---|---|
| 2.06 | -0.09 π | [1.80, 2.20] |
| **S1/S0 RMSE** | **S2/S0 RMSE** | **S3/S0 RMSE** |
| 0.29 | 0.31 | 0.31 |

Table 6. Fit results for the calculated results shown in Figure S13. Top row (left-to-right), optimal relative amplitude weighting ratio of $p_z/p_y$, the calculated relative phase shift correction $\phi(c_z$-$c_y)$, and interval in which the amplitude weighting reproduces a fit within 1% of the found optimal value. Bottom row: RMSE calculated from fitting our calculated S1/S0, S2/S0, and S3/S0 to the experimental results.

### 6.3.1 Experimental angle-resolved CL polarimetry (Figure 6c)

In this section we show the experimental- and calculated angle-resolved CL polarimetry results that are part of the data collected for the results shown in Figure 6c.

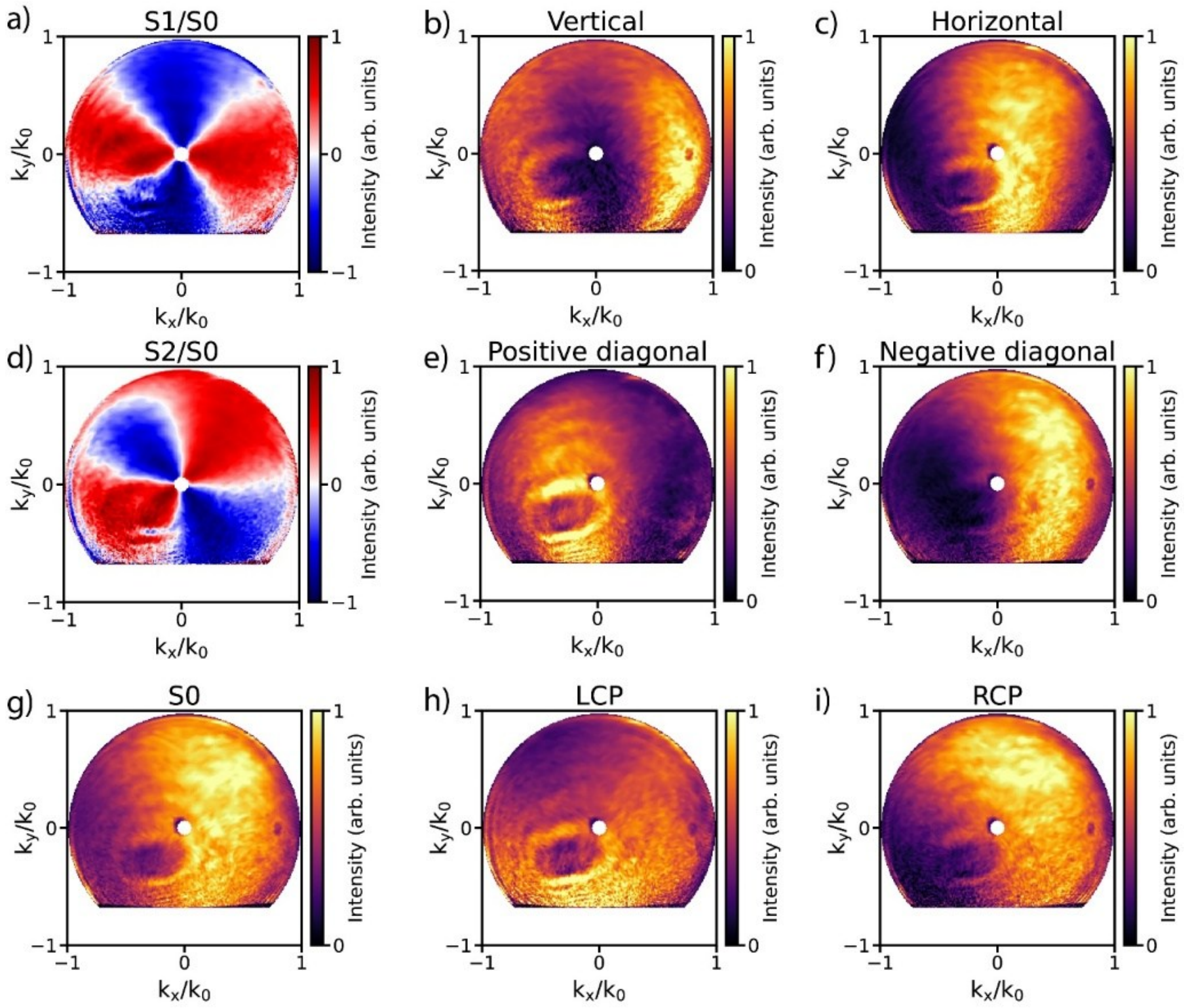

Figure S14. Experimental angle-resolved polarization-resolved CL emission (a) Normalized Stokes parameter S1/S0. (b,c) Angle-resolved linear vertically and horizontally polarized CL emission. d) Normalized Stokes parameter S2/S0. (e,f) Angle-resolved positive-diagonal (+45°) and negative-diagonal (−45°) polarized CL emission. (g) S0 parameter, used to normalize S1 to S3. (h,i) Angle-resolved left- and right handed circularly polarized CL emission.

### 6.3.2 Calculated angle-resolved CL polarimetry (Figure 6c)

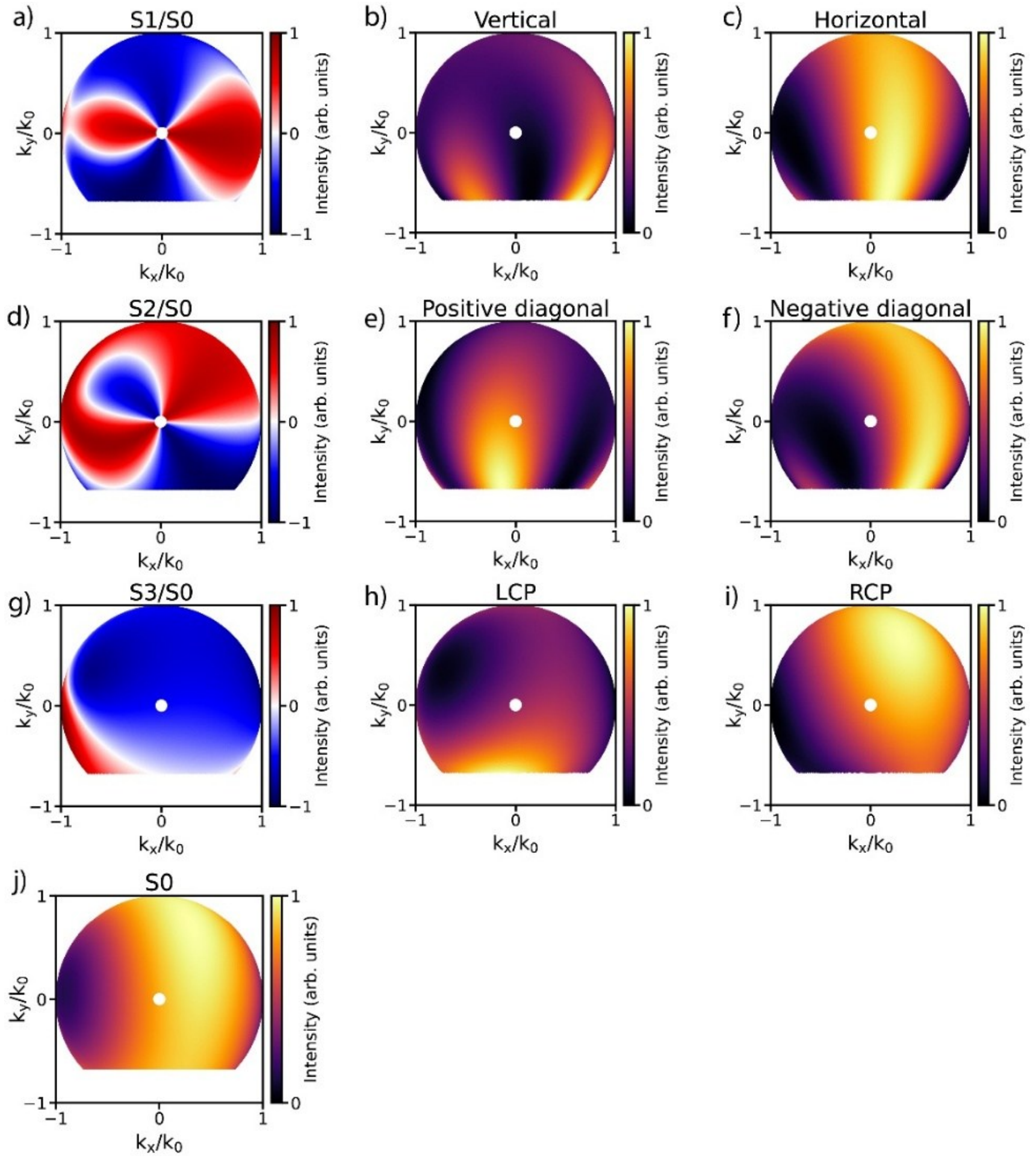

Figure S15. BEM-calculated angle-resolved polarization-resolved CL emission**.** (a) Normalized Stokes parameter S1/S0 (b,c) Angle-resolved vertically and horizontally polarized CL emission. (d) Normalized Stokes parameter S2/S0. (e,f) Angle-resolved positive-diagonal (+45°) and negative-diagonal (−45°) polarized CL emission. (g) Normalized Stokes parameter S3/S0. (h,i) Left- and right handed circularly polarized CL emission. (j) Stokes parameter S0, used for normalization

| **$\vert C_z/C_y \vert$** | **$\Delta\phi$ ($c_z$-$c_y$)** | **$C_z$ interval** |
|---|---|---|
| 2.76 | -0.06 π | [2.60, 2.80] |
| **S1/S0 RMSE** | **S2/S0 RMSE** | **S3/S0 RMSE** |
| 0.22 | 0.25 | 0.28 |

Table 7. Fit results for the calculated results shown in Figure S15. Top row (left-to-right), optimal relative amplitude weighting ratio of $p_z/p_y$, the calculated relative phase shift correction $\phi(c_z$-$c_y)$, and interval in which the amplitude weighting reproduces a fit within 1% of the found optimal value. Bottom row: RMSE calculated from fitting our calculated S1/S0, S2/S0, and S3/S0 to the experimental results.

### 6.4.1 Experimental angle-resolved CL polarimetry (Figure 6d)

In this section we show the experimental- and calculated angle-resolved CL polarimetry results that are part of the data collected for the results shown in Figure 6d.

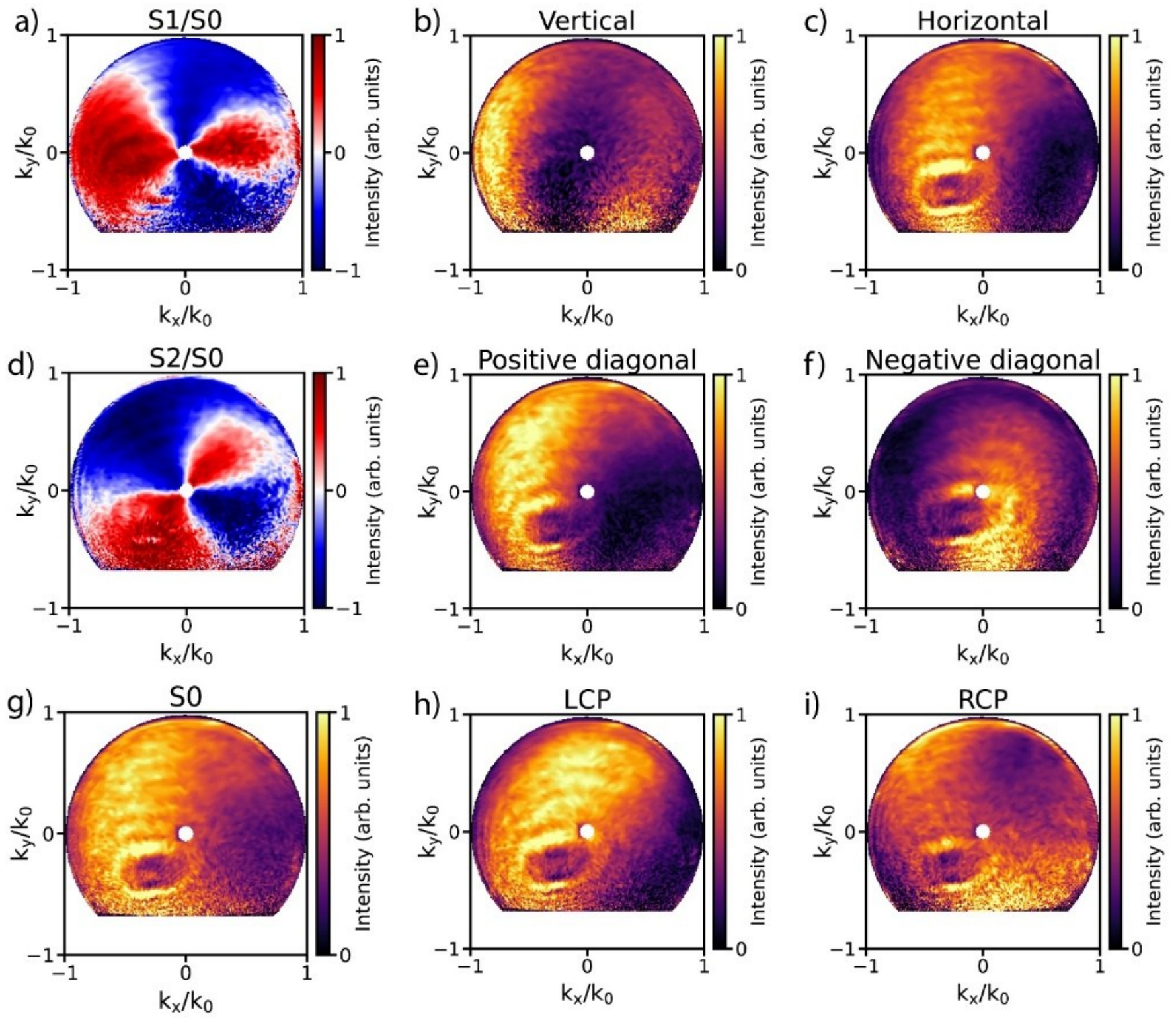

Figure S16. Experimental angle-resolved polarization-resolved CL emission (a) Normalized Stokes parameter S1/S0. (b,c) Angle-resolved linear vertically and horizontally polarized CL emission. d) Normalized Stokes parameter S2/S0. (e,f) Angle-resolved positive-diagonal (+45°) and negative-diagonal (−45°) polarized CL emission. (g) S0 parameter, used to normalize S1 to S3. (h,i) Angle-resolved left- and right handed circularly polarized CL emission.

### 6.4.2 Calculated angle-resolved CL polarimetry (Figure 6d)

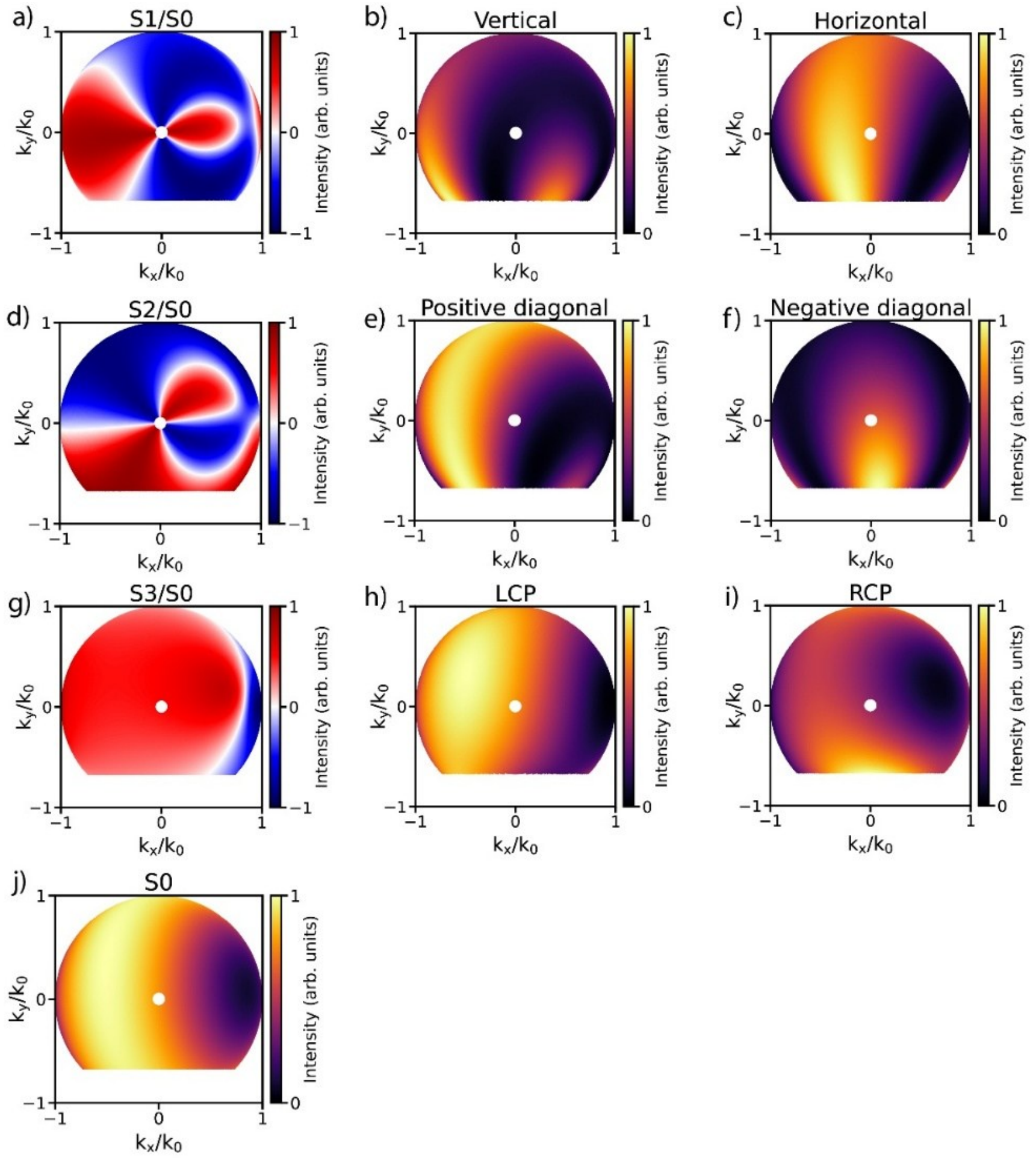

Figure S17. BEM-calculated angle-resolved polarization-resolved CL emission**.** (a) Normalized Stokes parameter S1/S0 (b,c) Angle-resolved vertically and horizontally polarized CL emission. (d) Normalized Stokes parameter S2/S0. (e,f) Angle-resolved positive-diagonal (+45°) and negative-diagonal (−45°) polarized CL emission. (g) Normalized Stokes parameter S3/S0. (h,i) Left- and right handed circularly polarized CL emission. (j) Stokes parameter S0, used for normalization

| $\lvert C_z/C_y \rvert$ | $\Delta\phi$ ($c_z$-$c_y$) | $C_z$ interval |
|---|---|---|
| 2.76 | -0.06 π | [2.60, 2.80] |
| **S1/S0 RMSE** | **S2/S0 RMSE** | **S3/S0 RMSE** |
| 0.22 | 0.25 | 0.28 |

Table 8. Parameters used to fit the simulations presented in Figure S4 to experimental results in Figure S3. Top row (left-to-right) relative amplitude weighting constants used for $p_x/p_z$, the relative phase difference, and the interval in which the amplitude weighting reproduces a fit close to our threshold. Bottom row: RMSE calculated from fitting our calculated S1/S0, S2/S0, and S3/S0 to the experimental results.